\documentclass[aip,jcp,reprint]{revtex4-1}
\usepackage{graphicx}
\usepackage{amsmath,amssymb}
\usepackage{subfig} 
\newcommand\abs[1]{\left|#1\right|}
\usepackage{xcolor}
\usepackage{chemformula}
\newcommand{\ba}{\begin{eqnarray}}
\newcommand{\ea}{\end{eqnarray}}
\newcommand{\be}{\begin{equation}}
\newcommand{\ee}{\end{equation}}
\usepackage{soul}
\begin{document}
	\title{Charge regulation of colloidal particles in aqueous solutions}
	\author{Amin Bakhshandeh}
	\email{amin.bakhshandeh@ufrgs.br}
	\affiliation{Instituto de F\'isica, Universidade Federal do Rio Grande do Sul, Caixa Postal 15051, CEP 91501-970, Porto Alegre, RS, Brazil}

	\author{Derek Frydel}
	\email{derek.frydel@usm.cl}
	\affiliation{Department of Chemistry, Federico Santa Maria Technical University, Campus San Joaquin,7820275,  Santiago,Chile}
	
	\author{Yan Levin } 
	\email{levin@if.ufrgs.br}
	\affiliation{Instituto de F\'isica, Universidade Federal do Rio Grande do Sul, Caixa Postal 15051, CEP 91501-970, Porto Alegre, RS, Brazil}
	\begin{abstract}
		We study charge regulation of colloidal particles inside aqueous electrolyte solutions. To stabilize colloidal suspension against precipitation, colloidal particles are synthesized with either acidic or basic groups on their 
		surface. In contact with water these surface groups undergo proton transfer reaction, resulting in colloidal surface charge.
		The charge is determined by the condition of local chemical equilibrium between hydronium ions inside the solution and at the colloidal surface.  We use a model of Baxter sticky spheres to explicitly calculate the equilibrium dissociation constants and to construct a theory which is able to quantitatively predict the effective charge of colloidal particles with either acidic or basic surface groups.  The predictions of the theory for the model are found to be in excellent agreement with the results of Monte Carlo simulations.  The theory is further extended to treat colloidal particles with a mixture of both acidic and basic surface groups. 
	\end{abstract}
	\maketitle
	\section{Introduction}
	Aqueous solutions are of great importance in biology and chemistry~\cite{ro1997,ANDELMAN1995,Messina_2009,Abrashkin}. In many cases
	such solutions are ionic. The long-range Coulomb interaction between charged particles is the mains source of difficulty for exploring the thermodynamics of such systems~\cite{levin,shen2017electrostatic,smith2016electrostatic,adar2018dielectric,Krishnan3,Frydeloettel,walker2011}.
	Because of organic functional groups many surfaces and membranes acquire surface charge when placed in water.  From their interaction with water and acid or base 
	these functional groups can either lose or gain a proton, becoming charged~\cite{alexander1984,trizac2002}. The amount of charge gained in this process
	depends on the pH of solution~\cite{adamson,Dawei} and the process is known as charge regulation (CR)~\cite{bakhshandeh2019,podgornik2018,frydel2019,avni2019charge,lind,majee2019,Yael2018,chan71white,pericet2004,VONGRUNBERG1999339,ozcelik2019electric,Podgornikexpansion,van2018solution,Markovich_2016,avni2019charge,Krishnan1,Krishnan2,
		Hartvig}.   CR is of great importance in colloidal science, biology, and chemistry~\cite{prieve1976,carnie1993,netz2002,Majee,hallett,lowen1,monica1,monica2,ZANDI2020,roshal2019ph,javidpour2019role,Zahler,PodgornikpH,Mikaellund,SHEN200592,Biesheuvel,Narahari,Burak,Kumar,Fleck,Lund2013,Grant2001,Elcock2001,Lund2005,Mason2008,Aguilar2010,Stahlberg1996,Tsao2000,Philippe}, and is responsible for the stability of many different systems~\cite{Markovich_2017,podgornik1991,leckband,Borkovecl2016,Maarten,Jyh-Ping,WOLTERINK200613,Ionel,Gong,Holger1999}.   
	
	The concept of charge regulation was first described by Linderstr{\o}m-Lang and later developed by many other 
	researchers~\cite{lind,kirkwood1952,marcus1955,lifson1957,PodgornikA2014,SMITH2020,safinya_h}. The first 
	quantitative implementation of charge regulation was done by Ninham and Parsegian (NP)~\cite{ninham1971} 
	who combined the idea of the local chemical equilibrium with the Poisson-Boltzmann theory introduced by Gouy 
	and Chapman sixty years earlier~\cite{chapman19,gouy1910}.
	The fundamental assumption of the NP theory is that the bulk association constants can be used to study proton transfer reactions with the surface adsorption sites.  Within the NP approach the bulk concentration of hydronium ions is replaced by the local density determined self consistently  by the Boltzmann distribution,
	%%%%%%%%%%%%%%%%%%%%%%%%%%%%%%
	\begin{equation}
	c_{\ch{H+}}^{surf}=c_{\ch{H_3O+}}^{bulk} \exp({-\beta \phi_0}),
	\label{Eq4}
	\end{equation} 
	%%%%%%%%%%%%%%%%%%%%%%%
	where  $\beta=1/k_BT$ and $\phi_0$ the electrostatic surface potential.
	
	The NP theory relies on Poisson-Boltzmann equation with the CR implemented as a new 
	boundary condition.  The two parameters that determine the boundary condition are the equilibrium 
	constant of the chemical reaction taking place at the surface, and the surface density of the chemical 
	groups.  Within the NP model, the surface is homogeneous, therefore, the model ignores the discrete structure 
	of surface chemical groups.  Another assumption is that the equilibrium constant is defined in terms of the concentrations of the reacting species rather than their activities. The validity of this assumption needs to be tested, since the concentration of hydronium ions can be quite large near a charged surface. 
	%Furthermore, the proton concentration 
	%in the equilibrium constant corresponds to the concentration at a contact with a 
	%surface.  (This implicitly assumes the sticky type of interactions between a proton and a surface). 
	Finally, 
	the value of the equilibrium constant at the surface
	is assumed to be the same as for the reaction in the bulk.  This is clearly  far from obvious.  
	% Alternatively, an equilibrium constant may be used as a fitting parameter.  
	
	In this paper we will focus on spherical colloidal particles with acidic and basic surface groups.   
	%The total colloidal charge is then renormalized  from its "bare" value -- when all the acidic surface groups are fully ionized  --  by the associated protons.  
	If a colloidal particle has $N_{site}$ basic functional groups on its surface 
	then the effective surface charge within the NP theory is found to be 
	%%%%%%%%%%%%%%%% equation %%%%%%%%%%%%%%%%%%%%%
	\begin{equation}    
	\sigma = \frac{ K_{Bulk}N_{site}q~c_a~\mathrm{e}^{-\beta \phi_0}}{4~\pi~(a+r_{ion})^2 (1 +  K_{Bulk}~c_a~\mathrm{e}^{-\beta\phi_0 }) },
	\label{Eq5}
	\end{equation} 
	where  $a$ is the colloidal radius  and $c_a$  is the bulk concentration of strong acid. 
	On the other hand if the surface has $N_{site}$ acidic groups, the effective surface charge is 
	%%%%%%%%%%%%%%%% equation %%%%%%%%%%%%%%%%%%%%%
	\begin{equation}    
	\sigma = -\frac{N_{site}~q}{4 \pi (a+r_{ion})^2}+\frac{ K_{Bulk}N_{site}q~c_a~\mathrm{e}^{-\beta \phi_0}}{4~\pi~(a+r_{ion})^2 (1 +  K_{Bulk}~c_a~\mathrm{e}^{-\beta\phi_0 }) },
	\label{Eq5a}
	\end{equation}
	%%%%%%%%%%%%%%%%%%%%	
	where $ K_{Bulk}$ is the bulk equilibrium association constant, which is the inverse of the acid dissociation constant $K_a$,  and $q$ is the elementary proton charge. The electrostatic surface potential $\phi_0$ must be calculated self-consistently by combining these expressions with the 
	mean-field Poisson-Boltzmann equation.

	The fundamental ingredient of the NP theory is the equilibrium constant.  
	In the original approach the equilibrium constant for the active sites on the colloidal surface was assumed to be the same as for the bulk solution, however, in the latter works the equilibrium constant was treated as a fitting parameter.  Clearly this is not very satisfactory, since it does not allow us to explicitly probe the validity of the theory. Although, the NP theory was a pioneering
	first step in understanding charge regulation in colloidal systems, in
	the absence of an explicit model on which the theory could be tested, the validity of the underlying approximations of the theory remains unclear. 
	A different approach was recently advocated by Bakhshandeh et al. in which a specific model of 
	of association was used to calculate exactly the bulk equilibrium constant for acid~\cite{bakhshandeh2019}.  The same  acidic groups where then placed on top of a spherical colloidal particle and the density profiles for hydronium cations and corresponding anions were calculated exactly  -- within this model -- using Monte Carlo simulations.  Knowledge of the exact equilibrium constant allowed us to explicitly compare the results of simulations with the NP theory.  It was found that NP approach deviated significantly from the predictions of simulations. For the specific case of acidic surface groups Ref. \cite{bakhshandeh2019} then introduced an alternative approach which was found to be in excellent agreement with the Monte Carlo simulations.  The objective of this paper is to extend the results of~\cite{bakhshandeh2019} to colloidal particles with basic surface groups, as well to the particles containing a mixtures of basic and acidic surface groups. 
	%{\color{red}{ We  will test the NP theory on models of association which 
	%can be solved exactly using computer simulation.  It is important to recall that NP approach is a mean-field theory.  On the other hand strong binding between hydronium ions and the surface charge can result in significant correlational effects.  It is precisely  the effect of these correlation on colloidal charge regulation that we want to explore in the present work.  }}
	
	We should note that the present theory applies directly only to the specific model of chemical association described below.  There are two levels of approximation that we use: 1 -- the microscopic model of acid-base association in terms of the Baxter sticky spheres, and 2 -- the approximations  used to theoretically solve the model. The advantage of this  two step approach is that the theory can be tested against an ``exact" solution of the microscopic model obtained using the computer simulations.   This allows us to separate the possible shortfalls of the theory from those of the microscopic model.   If the theory agrees with the ``exact" solution of the model, any shortfalls can then be attributed to the microscopic model of association and not to the approximations which had to be made to solve the model.  The  disadvantage of such
	approach is that the theory that we 
	develop applies only to the specific microscopic model of acid/base equilibrium and is not generally universal.   This, however, is the problem with any microscopic theory which does not explicitly take into account all the quantum effects associated with the charge transfer at the interface.  In the absence of such ``complete" theory, we expect that the approach advocated in the present paper will help to shed
	interesting new light on the mechanisms of charge regulation of nanoparticles and colloidal suspensions, and in particular on applicability of mean-field theories to study this intrinsically strong-coupling problem.   
	
	There are several possibilities for colloidal surface to acquire charge. The acidic functional groups, such as carboxyl \ch{COOH}, can become dissociated due to the following reaction
	\begin{equation}    
	\ch{HA} +\ch{H_2 O}   \rightleftarrows \ch{H_3 O+}  + \ch{A-},
	\label{Eq1}
	\end{equation} 
	resulting in a negatively charged surface.
	Alternatively, basic functional groups, which originally are not charged, can gain protons
	from hydronium ions and acquire a positive charge,
	\begin{equation}    
	\ch{B} +\ch{H_3 O+}   \rightleftarrows \ch{H_2 O}  + \ch{HB+}.
	\label{Eq2}
	\end{equation} 
	One example of such functional group is amine \ch{NH2}.

	The paper is organized as follows. In section II, we introduce a model of a colloidal particles with sticky sites and present the details of Monte Carlo simulations.
	In section III  we show how the equilibrium association constant can be calculated for sticky ions.  In section IV we present a model for a uniformly sticky colloidal particle.  In Section V this model is
	extended to account for discrete basic surface groups, and in Section VI and VII to discrete acidic groups.   In Section VIII we consider particles with a mixture of both basic and acidic surface groups and in Section IX we present our conclusions.

	\section{Theoretical background and Monte Carlo simulations details}
	\subsection{Theoretical background} 
	%A primary component of the microscopic model used in this
	%work are the sticky spheres together with a chemical interpretation
	%according to which two sticky spheres in direct contact are regarded as chemically bonded forming a new molecule.
	
	To study charge regulation of a colloidal surface we use a model of Baxter sticky spheres~\cite{baxter,frydel2019,frydel2019_2}. The sticky potential was previously used to study  gelation in globular proteins~\cite{Frenkel2003,Foffi2005}, chemical association in weak acid-base reactions~\cite{Herrera}, and sticky-charged wall model~\cite{Blum,Huckaby,Jeffery,Greathouse}.
	In our model, sticky interactions take on a physical
	interpretation of a chemical bond between proton and
	acid/base groups.  This is not the first time that a sticky interaction is used to model a chemical bond, to capture some aspect of quantum mechanics in an otherwise classical description. The idea has been around for some time and
	reaches back to 1980  in particular, the 
	work of Blum and Herrera~\cite{Blum,Herrera}, and Werthaim~\cite{Wertheim1986} for directional chemical bonding. Sticky interactions
	continue to this day being an important part of soft-matter modeling~\cite{Wang2012}. In the present work  sticky interactions,
	and their quantum-chemical interpretation, will be used to study charge regulation of nanoparticles with surface acid and base groups.  The results obtained, therefore, are only valid within the specific microscopic model.  
	The model, of course, can be extended and
	modified to represent a different charge regulated system. For example, directional chemical bonding could
	be introduced by making a sphere sticky in limited regions. The size and shape of an absorbing molecule could
	be changed. These alternatives are not explored in the present
	work.
	
\begin{figure}[h] 
		\begin{center}
			\begin{tabular}{rrr}
				\includegraphics[height=0.22\textwidth,width=0.22\textwidth]{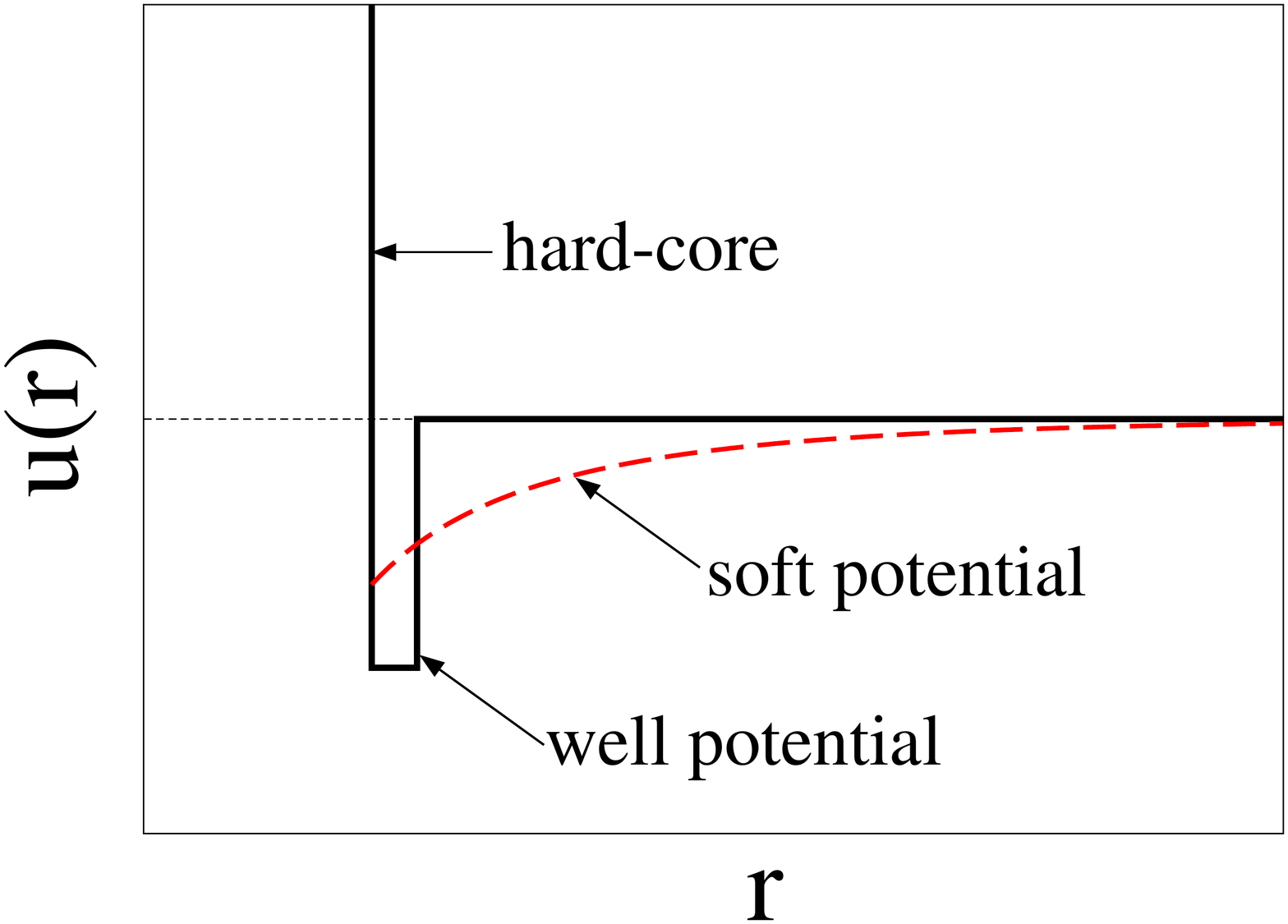}&
				\includegraphics[height=0.22\textwidth,width=0.22\textwidth]{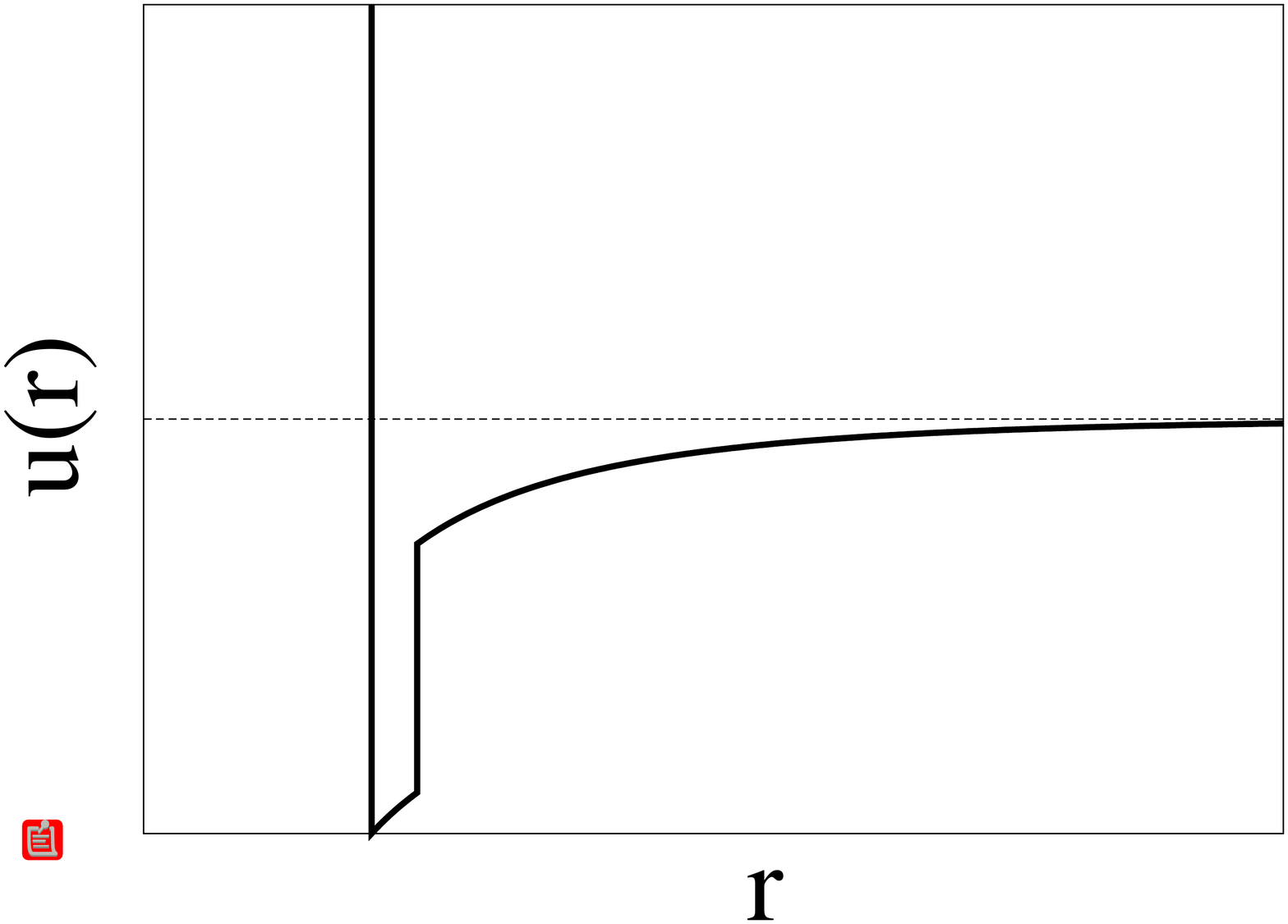}\\
			\end{tabular}
		\end{center}
		\caption{Schematic representation of the sticky-hard-sphere potential plus soft interaction.  The 
			sticky part is represented as a narrow well potential;  (a) depicts hard-core plus attractive well 
			and soft potential separately, and (b) as a combination.  An important observation is that the soft interaction is 
			active within the attractive well.  }
		\label{fig:u_well} 
	\end{figure}

	The hydronium ion can become adsorbed on an acidic or basic functional group to form a molecule~\ch{H^+A^-} or~\ch{H^+B}, respectively.   
	To model the binding  between \ch{H+} and \ch{A^-} or  \ch{B} we use an attractive square well potential with a repulsive hard core~\cite{frydel2019,bakhshandeh2019}. 
	The first component of the interaction potential is the hard-core repulsion,
	\begin{equation}  
	u_{hs}(r) =
	\begin{cases}
	\infty, & r<d,\\
	0, &        r > d,
	\end{cases} 
	\label{eq:u_hs}
	\end{equation}
	where $d$ is the diameter of particles.  The second component is a narrow attractive well,
	\begin{equation}  
	u_{well}(r) =
	\begin{cases}
	0, & r<d,\\
	-\varepsilon, &  d<r< d+\Delta,\\
	0, &        r > d+\Delta.
	\end{cases} 
	\label{eq:u_well}
	\end{equation}
	To generalize the model, we also include a soft potential $u_{sf}(r>d)$, so that 
	the total pair potential becomes $u_{tot} = u_{hs} + u_{well} + u_{sf}$,  
	\be  
	u_{tot}(r) =
	\begin{cases}
		\infty, & r<d,\\
		-\varepsilon + u_{sf}(r), &  d<r< d+\Delta,\\
		u_{sf}(r), &        r > d + \Delta.
	\end{cases} 
	\label{eq:u_tot}
	\ee
	We note that the soft potential $u_{sf}(r)$ is effective from $r=d$, 
	as illustrated in Fig. (\ref{fig:u_well}).  
	%Whether or not a soft potential is active inside a well region may 
	%appear as an irrelevant detail in the sticky limit $\Delta\to 0$.  
	%But as it will be demonstrated, different constructions will lead to different expressions. 
	%%%%%%%%%%%%%%%%%%%%%%
	\graphicspath{{figures/}}
	For illustration, we consider a very simple scenario comprised of two particles interacting via a 
	pair potential in Eq. (\ref{eq:u_tot}) and confined to a spherical region of radius $R$.  To make 
	the demonstration even simpler, one particle is fixed at the origin and only a second particle is free.  The 
	resulting partition function has two parts, 
	\begin{eqnarray}
	\cal{Z} &&= 4\pi\int_d^{R} r^2 e^{-\beta [u_{well}(r)+u_{sf}(r)]}  dr = \nonumber \\
	&& 4\pi  e^{\beta \varepsilon} \int_d^{d+\Delta}  r^2 e^{-\beta u_{sf}(r)}  dr + 4\pi \int_{d+\Delta}^{R} r^2 
	e^{-\beta u_{sf}(r)}  dr. 
	\label{eq:Z1}
	\end{eqnarray}
	Assuming a small $\Delta$, we can expand $\cal{Z}$ in $\Delta$, yielding 
	\ba
	\cal{Z} &=& 4\pi  d^2 e^{\beta \varepsilon} e^{-\beta u_{sf}(d)} \bigg[\Delta\bigg(1- e^{-\beta \varepsilon}\bigg) \nonumber\\
	&& ~ + ~ \bigg(\frac{1}{d} - \frac{1}{2}\frac{d\beta u_{sf}(r)}{dr}\bigg|_{r=d} \bigg)\Delta^2 + \dots \bigg] 
	\nonumber\\ && ~ + ~
	4\pi \int_{d}^{R} r^2 e^{-\beta u_{sf}(r)}  dr.  
	\ea
	In the limit $\Delta\to 0$ only the last term does not vanish.  However, if, at the same time as 
	$\Delta\to 0$, $\varepsilon\to \infty$, then some of the expansion terms must also be retained.  
	The correct way to carry out the limit is to require that $\Delta e^{\beta \varepsilon}=\rm{const}$,  
	which is referred to as the Baxter limit \cite{baxter}, and yields 
	\be
	{\lim_{\substack{\Delta\to 0 \\  \varepsilon\to\infty}}} {\cal{Z}} = 4\pi  d^2 l_g e^{-\beta u_{sf}(d)} 
	+ 4\pi \int_{d}^{R} r^2 e^{-\beta u_{sf}(r)}  dr,
	\label{eq:Z2}
	\ee
	where we introduced the ``sticky length" defined as 
	\be
	l_g = \lim_{\substack{\varepsilon\to\infty \\ \Delta\to 0}} \Delta e^{\beta \varepsilon}.  
	\label{eq:lg}
	\ee
	
	%If, alternatively, we assume that $u_{sf}(r)$ is active in the region $r>d+\Delta$, the partition function becomes 
	%\be
	%Z' = 4\pi  e^{\beta \varepsilon} \int_d^{d+\Delta}  r^2 dr + 4\pi \int_{d+\Delta}^{R} r^2 e^{-\beta u_{sf}(r)}  dr,
	%\label{eq:Z'1}
	%\ee
	%which in the Baxter limit yields 
	%\be
	%\lim_{\substack{\Delta\to 0 \\ e^{\beta \varepsilon}\to\infty}} Z' = 4\pi d^2 l_g + 4\pi \int_{d}^{R} e^{-\beta u_{sf}(r)} r^2 dr.  
	%\label{eq:Z'2}
	%\ee
	%This construction yields a different first term from that in Eq. (\ref{eq:Z2}).  
	
	%To intuitively understand the difference between the two constructions, we consider charged particles. 
	%For example, if sticky interactions are between oppositely charged particles, it is natural to expect 
	%that more particles would be found in paired configurations.  This behavior is captured by the first 
	%construction but not the second one.  Consequently, our model of charge regulation follows
	%the first construction of sticky interactions.  
	
	Of great concern for simulations is the width $\Delta$ of the well potential, since in practice the  
	exact Baxter limit cannot be attained  and $\Delta$ must remain finite.  
	To estimate what is sufficiently small value of $\Delta$, 
	we consider the previous simple system with $u_{sf}=0$, for which the exact partition function is
	\be
	{\cal{Z}} = 4\pi d^2 l_g\bigg[1 + \frac{\Delta}{d} + \frac{1}{3}\bigg(\frac{\Delta}{d}\bigg)^2\bigg]  
	+ \frac{4\pi R^3}{3} \bigg[1 - \bigg(\frac{d + \Delta}{R}\bigg)^3\bigg].  
	\ee
	If we ignore the second term in square brackets, assuming $R\gg d$, we conclude that the well potential becomes sticky if 
	$\Delta/d\ll 1$.  In practice, we find that $\Delta/d \approx 0.01$ is sufficiently small to suppress most 
	contributions of finite $\Delta$.

	The Baxter sticky potential may appear analogous to a delta function potential often used in quantum mechanics.  
	This, however, is misleading.  The well potential in Eq. (\ref{eq:u_well}) transforms into the delta function 
	in the limits $\Delta\to 0$ and $\varepsilon\to\infty$, while the product $\Delta \varepsilon$ is held fixed.   On the other hand, the Baxter limit, requires that $\Delta e^{\beta \varepsilon}$ remains constant.  To see this more clearly~\cite{frydel2019} 
	we define  
	%%%%%%%%%%%%%%%% equation %%%%%%%%%%%%%%%%%%%%%
	\begin{equation}  
	f(r) =
	\begin{cases}
	\frac{1}{\Delta}, &        d \le r \le d+\Delta,\\
	0, &         r< d~ \text{or }~r>d+ \Delta.
	\end{cases} 
	\label{Delta}  
	\end{equation} 
	The Boltzmann factor then can be written as
	%%%%%%%%%%%%%%%% equation %%%%%%%%%%%%%%%%%%%%%
	\begin{equation}  
	e^{-\beta u_{well}(r)} = 1+\Delta(\mathrm{e}^{\beta \epsilon} -1)  f(r), 
	\label{Bol}
	\end{equation} 
	which in the Baxter limit reduces to
	\be
	\lim_{\substack{\varepsilon \to\infty \\ \Delta\to 0}}e^{-\beta u_{well}(r)} = 
	1 + l_g\delta(r-d),   
	\label{eq:Bfactor}
	\ee
	with the sticky length given by $l_g \equiv \Delta(\mathrm{e}^{ \beta \epsilon} -1)$.  In the Baxter limit
	the $-1$ in the definition of $l_g$ can be neglected, however, in the simulations with finite $\Delta$ we will use the exact expression for $l_g$.  The 
	sticky potential itself is then
	\be
	\beta u_{well}(r) = -\ln \bigg[1 + l_g  \delta(r-d)\bigg],  
	\ee
	which shows that  it is weaker than the delta function potential.  Indeed, a delta function potential would result in an irreversible association between the sticky spheres.  Finally, we note that if the expression (\ref{eq:Bfactor}) is used in the partition function Eq. (\ref{eq:Z1}), we will arrive directly at the Eq. (\ref{eq:Z2}).

	\subsection{ Monte Carlo simulations details}
	
	We are now in a position to implement numerical simulations.   The simulations are performed inside a spherical Wigner-Seitz (WS) cell of radius $R$.  A spherical colloidal particle of radius  $a$ is placed in the center of the WS cell.  The radius of the cell is determined by the colloidal volume fraction of the suspension, $\varphi_c=a^3/R^3$.   The motivation for using WS is that  for small salt concentration colloidal system may crystallize, in which case thermodynamics will be very well described by the WS cell model, with a Donnan potential used to control the charge neutrality.  In fact, even for the disordered state the WS approach to thermodynamics is found to lead to osmotic pressures in excellent agreement with experiments~\cite{tamashirodonnan}.  In a sense, the many body colloid-colloid interactions in the grand-canonical ensemble are all included through the boundary condition of vanishing electric field at the WS cell boundary. 
	
\begin{figure}
		\begin{center}
			\includegraphics[width=7cm]{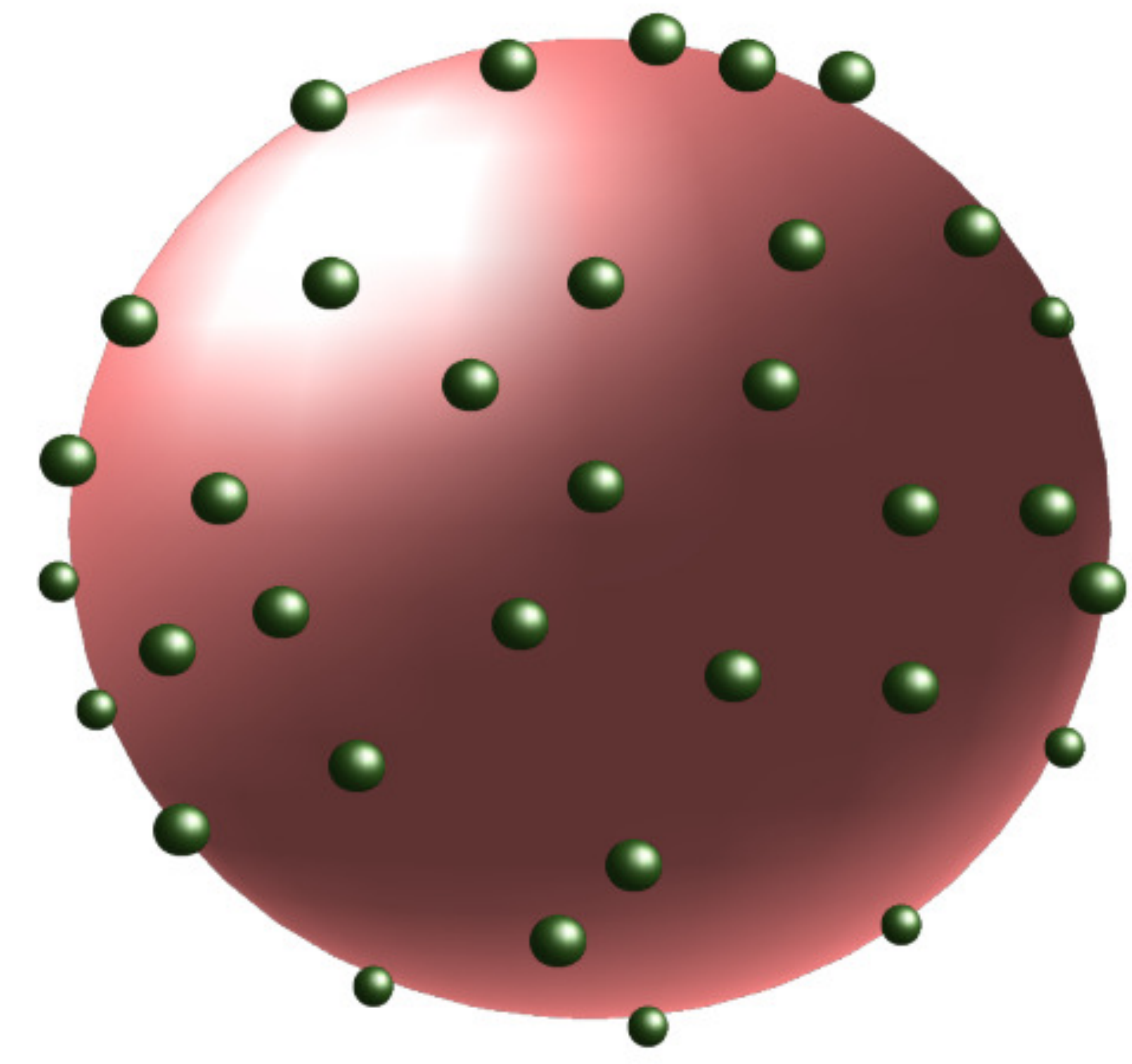}
		\end{center}
		\caption{The Baxter's sticky spherical sites on the colloidal surface.  }  
		\label{figmodel}
	\end{figure}

	The colloidal particle has $N_{site}$ adsorption sites randomly distributed over its surface.  Each adsorption site is a sphere of diameter $d$, see Fig. \ref{figmodel}.  If an active site is basic -- has zero charge --- it interacts with the hydronium ions through the hard core and the Baxter sticky potential, Eq.(\ref{eq:u_tot}).  On the other had if the site is acidic --- has charge $-q$, where $q$ is the proton charge --- in addition to the Baxter and hard core interactions, there is also a long range Coulomb potential between the adsorption site and all the ions inside simulation cell. In this work
	all the ions and the adsorption sites have diameter $4$ \AA~ and the colloidal particle has radius of $a=100$ \AA. 
	
	The system is connected to a reservoir of strong acid at concentration $10^{-pH}$,  and a reservoir of 1:1 strong electrolyte at concentration $c_s$.
	%%%%%%%%%%%%%%%%%
	%%%%%%%%%%%%%%%%%%%%%
	The solvent is considered to be a uniform dielectric of permittivity $\epsilon_w = 80 \epsilon_0$ and the Bjerrum length is $\lambda_B=q^2/\epsilon_w k_B T = 7.2 $\AA. 
	The total interaction potential is 
	\begin{equation}
	\begin{split}
	U =  \sum_{i>j}\frac{  q_i~q_j }{\epsilon_w |\bold{r}_i-\bold{r}_{j}|}+\sum{}^{'} u_{well}(\bold{r}_i),
	\end{split}
	\label{Eqadded_1}
	\end{equation}
	where the first sum is over all the charged particles, including the adsorption sites, and the second sum is for the sticky interaction between the hydronium ions and the adsorption sites. The hardcore interaction between ions, sites, and colloidal surface is implicit.   The restriction on the second sum indicated by the prime is due to the fact that each functional site can adsorb at most one hydronium ion. This is the case for carboxyl or amine groups. Therefore, once there is a hydronium ion within the distance $\Delta$ of the adsorption site, the short range sticky potential of this site with other hydronium ions is switched off.  In this paper we will not consider more complicated metal oxide ions which can adsorb more than one proton. To perform simulations we used Metropolis algorithm~\cite{metropolis}. For large WS cells, when a system establishes a well defined bulk concentration far from the colloidal surface,  we can use canonical Monte Carlo simulations~\cite{Frenkel}.  For large colloidal volume fractions, when WS is small and bulk concentration is not reached inside the cell, we use the grand canonical  Monte Carlo simulations~\cite{Frenkel}.  This is done in order to have a well defined reservoir  concentrations of acid $c_a$ and salt $c_s$, which are necessary to compare the theory with the simulations.  In both types of simulations we have used  $5 \times 10^6$ MC steps for equilibration and $10^4$ steps for production. 
	
\begin{figure}
		\begin{center}
			\includegraphics[width=9cm]{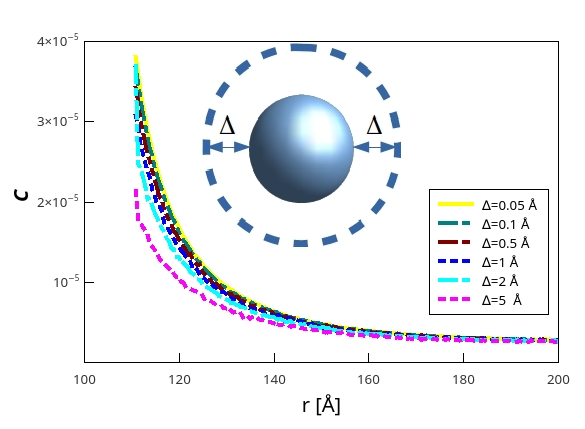}
		\end{center}
		\caption{The density profiles of ions for different size $\Delta$ and fixed $l_g$=109.9\AA. The density is plotted in terms of the number of particles per \AA$^3$.  }  
		\label{conv}
	\end{figure}

	We first check the convergence of MC results to the Baxter sticky limit by studying systems with different values of  $\Delta$ and $\epsilon$, while keeping fixed the sticky length $l_g$. Fig.~\ref{conv}, shows the rapid convergence to the Baxter limit, with decreasing value of $\Delta$.
	%%%%%%%%%%%%% end of figure %%%%%%%%%%%%%%%%%
	\section{Equilibrium constant for particles interacting via a pair potential}
	
	To connect the simulations presented in the previous section with the NP theory we must relate
	the sticky length with the bulk association constant.

	\subsection{Neutral pairs}

\begin{figure}[h] 
		\begin{center}
			\begin{tabular}{rrrr}
				\includegraphics[width=9cm]{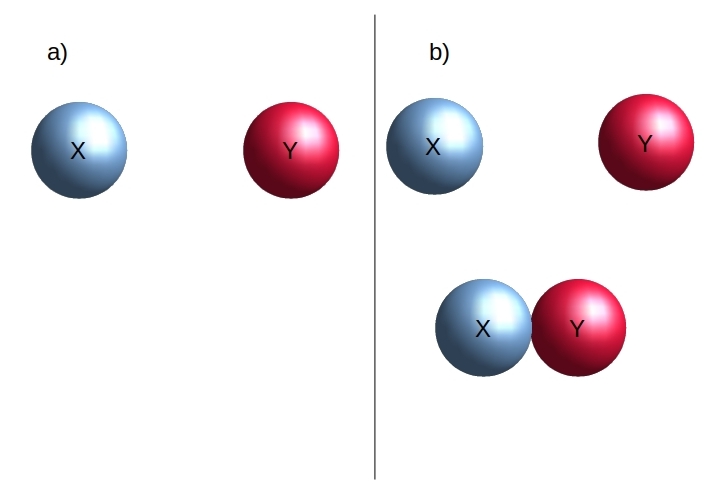}
			\end{tabular}
		\end{center}
		\caption{The two representations correspond to a) a physical and b) a chemical 
			interpretation. }
		\label{fig:config} 
	\end{figure}

	We first consider a general two-component system of sticky spheres.  To avoid confusion with 
	previous labels, we designate the ``atoms" of each species as \ch{X} and \ch{Y}.  The 
	interaction between atoms of the same species is
	\be
	u_{xx}(r) = u_{yy}(r) = u_{hs}(r) + u_{sf1}(r),
	\label{eq:uxx}
	\ee
	and between the atoms of different species is
	\be
	u_{xy}(r) = u_{hs}(r) + u_{sf2}(r) + u_{well}(r),
	\label{eq:uxy}
	\ee
	This is the, so called, ``physical picture", in which only atoms exist. Alternatively, we can
	regard two atoms $X$ and $Y$ in contact to form a molecule $XY$.  This corresponds to the 
	``chemical picture", see Fig. (\ref{fig:config}) for illustration.  In the chemical picture, we have free
	atoms $X$ and $Y$, and molecules $XY$ which are in ``chemical" equilibrium~\cite{hill}, 
	%%%%%%%%%%%%%%%%%%%
	\be
	\ch{X} +\ch{Y}   \rightleftharpoons \ch{XY}.
	\label{eq:reaction_xy}
	\ee
	%%%%%%%%%%%%%%%%%%%%%%%%
	At most two atoms \ch{X} and \ch{Y} are permitted to interact via a sticky potential. 
	Without this restriction, one has to account for the presence of triplets \ch{XYX}, quartets 
	\ch{XYXY}, and other higher order formations, together with their corresponding chemical 
	reactions.  
	%%%%%%%%%%%%%%%%%%%%%%
	\graphicspath{{figures/}}
	%%%%%%%%%%%%%%%%%%%%%%%
	
	To obtain the equilibrium constant for the chemical reaction in Eq. (\ref{eq:reaction_xy}) we
	compare the equations of state calculated using the physical and the chemical pictures.  Clearly
	the osmotic pressure calculated using the two interpretations of the same physical reality has to be same. 
	
	Within the physical interpretation, the system is comprised of two types of atoms, \ch{X} and \ch{Y}, 
	and the formation of pairs \ch{XY} is devoid of any special meaning.  
	The virial expansion of the osmotic 
	pressure up to second order in bulk concentration $c_i$ is~\cite{mcquarrie}
	\be 
	\beta P_{phys} = c_{x} +c_{y} +B_{xx}c_{x}^2 +B_{yy}c_{y}^2+2 B_{xy} c_{x} c_{y} + \dots, 
	\label{eq:P_phys}
	\ee
	where $B_{ij}$ are the second virial coefficients defined as 
	\be
	B_{ij} = 2 \pi  \int_{0}^{\infty} \Big(1-\mathrm{e}^{- \beta u_{ij}(r)}\Big) \, r^2 dr, 
	\label{Eq3-2}
	\ee
	and whose various contributions are
	\ba
	&& ~~~~~ ~~~~~ ~~ B_{xx} = B_{yy} = B_{hs} + B_{sf1}, \nonumber\\
	&& ~~~~~ ~~~~~ ~~ B_{xy} = B_{hs} + B_{st} + B_{sf2} ,
	\ea
	which, after evaluation, become 
	\ba
	&& ~~~~~ ~~~~~  ~~ B_{hs} = \frac{2\pi d^3}{3}, \nonumber\\
	&& ~~~~~ ~~~~~  ~~ B_{st} = -2 \pi l_g d^2 e^{-\beta u_{sf2}(d)}, \nonumber\\
	&& ~~~~~ ~~~~~  ~~ B_{sf1} = 2 \pi  \int_{d}^{\infty} \Big(1-\mathrm{e}^{- \beta u_{sf1}(r)}\Big) \, r^2 dr, \nonumber\\
	&& ~~~~~ ~~~~~  ~~ B_{sf2} = 2 \pi  \int_{d}^{\infty} \Big(1-\mathrm{e}^{- \beta u_{sf2}(r)}\Big) \, r^2 dr, \nonumber\\
	\label{eq:B}
	\ea
	where $B_{st}$ was evaluated using the Boltzmann factor in Eq. (\ref{eq:Bfactor}).  
	Inserting these contributions into the expansion in Eq. (\ref{eq:P_phys}) yields 
	\ba
	\beta P_{phys} &=& c_{x} +c_{y} \nonumber\\
	&+& B_{hs} \big(c_{x} + c_{y}\big)^2 + B_{sf1}\big(c_{x}^2 + c_{y}^2\big) + 2B_{sf2}c_xc_y \nonumber\\
	&+& 2 B_{st} c_{x} c_{y} + \dots, 
	\label{eq:P_phys2}
	\ea
	where the first line is the ideal-gas contribution, the second line is the second order correction due to hard-core and 
	soft interactions, and the third line is the second order correction due to the sticky potential.  
	
	To formulate the equation of state within the chemical picture, we need to define the 
	concentrations of free atoms \ch{X}, \ch{Y}, and of molecules \ch{XY}, designated by the superscript *:
	\ba
	&& ~~~~~~~~ ~~~~~~~~ ~~~~~~~ c^*_{x} =c_{x} - c^*_{xy}, \nonumber\\
	&& ~~~~~~~~ ~~~~~~~~ ~~~~~~~ c^*_{y }=c_{y} - c^*_{xy}, \nonumber\\
	&& ~~~~~~~~ ~~~~~~~~ ~~~~~~~ c^*_{xy} = K_{Bulk}{c_{x}^*} c_{y}^*, 
	\label{eq:cxcy}
	\ea
	where the last equation was obtained using the definition of the equilibrium constant
	of the reaction in Eq. (\ref{eq:reaction_xy}), 
	\be
	K_{Bulk} = \frac{c^*_{xy}}{c^*_{x}c^*_{y}}, 
	\label{eq:Keq}
	\ee  
	valid in the dilute limit where activities are approximated by concentrations.  
	To second order in concentrations $c_{i}$, Eq. (\ref{eq:cxcy}) can be written as 
	\ba
	&& ~~~~~~~~ ~~~~~~~~  c^*_{x} =c_{x} - K_{Bulk}{c_{x}} c_{y} + \dots, \nonumber\\
	&& ~~~~~~~~ ~~~~~~~~  c^*_{y }=c_{y} - K_{Bulk}{c_{x}} c_{y} + \dots, \nonumber\\ 	
	&& ~~~~~~~~ ~~~~~~~~  c^*_{xy} = K_{Bulk}{c_{x}} c_{y} + \dots.
	\label{eq:cxcy2}
	\ea
	In the chemical picture, the interactions between free atoms \ch{X} and \ch{Y} do not include 
	sticky interaction, 
	$$
	u'_{xy}(r) = u_{hs}(r) + u_{sf2}(r).  
	$$
	which acts only within a molecule \ch{XY}.   
	Without the sticky interaction, the modified second virial coefficient in the chemical interpretation is 
	\be
	B'_{xy} = B_{hs} + B_{sf2}.  
	\ee
	
	The osmotic pressure up to second order in concentrations $c_i$ is then
	%%%%%%%%%%%%%%%% equation %%%%%%%%%%%%%%%%%%%%%
	\be 
	\beta P_{chem} = c_{x}^* + c_{y}^* + c_{xy}^* + B_{xx}c_{x}^{*2} + B_{yy}c_{y}^{*2} + 2 B'_{xy} c_{x}^* c_{y}^* + \dots.  
	\label{eq:P_chem}
	\ee
	The terms 
	\be
	2B_{x,xy}c_{x}^*c_{xy}^{*} + 2B_{y,xy}c_{y}^*c_{xy}^{*} + B_{xy,xy}c_{xy}^{*2}, 
	\ee
	that are second order in $c_{i}^*$ 
	%proportional to $c^*_{x}c^*_{xy}$, $c^*_{y}c^*_{xy}$, and $c^*_{xy}c^*_{xy}$
	are omitted since, due to  $c_{xy}^*\approx K_{Bulk}{c_{x}} c_{y}$ in Eq. (\ref{eq:cxcy2}), 
	they are of higher order in $c_i$.  
	Using formulas in Eq. (\ref{eq:cxcy2}), and substituting for the coefficients $B_{ij}$, Eq. (\ref{eq:P_chem}) 
	becomes 
	\ba
	\beta P_{chem} &=& c_{x} +c_{y} \nonumber\\
	&+& B_{hs} \big(c_{x} + c_{y}\big)^2 + B_{sf1}\big(c_{x}^2 + c_{y}^2\big) + 2B_{sf2}c_xc_y \nonumber\\
	&-& K_{Bulk} c_{x} c_{y} + \dots, 
	\label{eq:P_chem2}
	\ea
	Setting $P_{phys}=P_{chem}$, and matching the terms of the same order yields 
	%Comparing this result with Eq. (\ref{eq:P_phys2}) yields 
	%the equilibrium constant, 
	%%%%%%%%%%%%%%%%%%%%%%%%%%%%%%%%% 
	\begin{equation}  
	K_{Bulk}  = -2B_{st} = 4\pi l_g {d}^{2} e^{-\beta u_{sf2}(d)},
	\label{eq:Keq2}
	\end{equation}
	%%%%%%%%%%%%%%%%%%%%%%%%%%%%%%%%%%
	
	We note that the above derivation assumes a dilute limit, where the definition of $K_{Bulk}$ in Eq. (\ref{eq:Keq}) 
	and the second order expansion of $\beta P$ are valid.  The result in Eq. (\ref{eq:Keq2}), however, is exact 
	for any concentration.  This is because the quantity $K_{Bulk}$ itself is independent of concentrations.  
	We simply took advantage of this fact and chose the limit where all the expressions are the simplest. 
	
	If we set $u_{fs2} = 0$ , the  equilibrium constant becomes  
	%%%%%%%%%%%%%%%%%%%%%%%%%%%%%%%%% 
	\begin{equation}  
	K_{Bulk}  =  4\pi l_g {d}^{2},
	\label{Eq3-5}
	\end{equation}
	%%%%%%%%%%%%%%%%%%%%%%%%%%%%%%%%%%
	which is appropriate for the pair formation between bases and hydronium ions.  On the other hand, as we will see in the following section,  Eq. (\ref{eq:Keq2}) with $u_{sf2}$ corresponding to the Coulomb potential will be appropriate for weak acid-hydronium equilibrium constant.
	
	\subsection{Charged pairs}
	
	In bulk, acid ``molecule" dissociates resulting in a hydronium ion and a corresponding anion:  
	%%%%%%%%%%%%%%%%%%%
	\begin{equation}    
	\ch{HA}  \rightleftarrows   \ch{H^+} +\ch{A^-}  ,
	\label{Eqr}
	\end{equation} 
	%%%%%%%%%%%%%%%%%%%%%%%%
	The thermodynamics of bulk electrolytes, even without covalent bonding between the ions, is complicated by the  
	divergence  of the virial expansion due to the long range nature of the Coulomb interaction. Instead a certain class of perturbative diagrams must be summed together
	to obtain a finite result~\cite{mcquarrie}.  This leads to a non-analytic term in the density expansion of the osmotic pressure 
	which scales with electrolyte concentration as $c^{3/2}$.  The next order term which scales as $c^2$ can be interpreted as the result of Bjerrum anion-cation pair formation.   
	In the case of purely electrostatic interactions, the equilibrium constant for such cluster formation was derived by  Ebeling~\cite{eblingo68} considering the
	exact density expansion of the equation of state up to  $\mathcal{O} \left(c^{5/2}\right)$~\cite{Falkenhagen}.
	The Ebeling equilibrium constant is:
	\begin{equation}  
	\begin{split}  
	K_{Eb} = 8\pi d^3\left\{ \frac{1}{12} b^3 \right. 
	\hspace*{0cm}  \left[\mathrm{Ei}\left(b\right) 
	-\mathrm{Ei}\left(-b\right)\right]-\frac{1}{3}\cosh{b}-\\
	\frac{1}{6}b\sinh{b}-
	\left.  \frac{1}{6}b^2\cosh{b}+\frac{1}{3}+\frac{1}{2}b^2\right\},  
	\end{split}  
	\label{Eqa1}
	\end{equation}
	where $b =\frac{\lambda_B}{d}$.
	For  large values of $b$ (strong coupling limit), the equilibrium constant can be expanded asymptotically to give:   
	\begin{equation}  
	K_{Eb} = 4 \pi a^3 \frac{\mathrm{e}^b}{b} \left(1+\frac{4}{b}+\frac{4\times 5}{b^2}+\frac{4\times 5 \times 6}{b^3}+...\right) \,.
	\label{Eqa2}
	\end{equation}
	This may be compared with the Bjerrum phenomenological association constant for formation of anion-cation pairs
	\begin{equation}
	K_{Bj} = 4\pi \int_{d}^{R_{Bj}}\mathrm{e}^{ \frac{\lambda_B}{r}}r^2 dr, 
	\end{equation}
	where $R_{Bj}=\lambda_B/2$ is the Bjerrum cutoff.    
	In the strong coupling limit (low temperatures), $K_{Bj}$ is completely insensitive to the precise value of cutoff $R_{Bj}$~\cite{levin}. Furthermore, the low temperature expansions for $K_{Eb}$ and $K_{Bj}$ are found to be 
	identical~\cite{levin}.  One can then interpret the Ebeling equilibrium constant as the analytic continuation of $K_{Bj}$ over the full temperature range. With this observation it becomes easy to obtain the 
	equilibrium constant for sticky electrolytes.  In the spirit of Bjerrum, we then write
	\begin{equation}  
	K_{Bulk} = 4\pi \int_{d}^{R_{Bj}}\mathrm{e}^{-\beta u_{st}(r)+ \frac{\lambda_B}{r}} r^2 dr.
	\label{Eq4a}
	\end{equation}
	Using Eq.(\ref{eq:Bfactor}) we obtain
	\begin{equation}  
	K_{Bulk} = 4\pi \int_{d}^{R_{Bj}}\left[(1+l_g \delta(r-d)\right]\mathrm{e}^{ \frac{\lambda_B}{r}}r^2 dr\,,
	\label{Eq17}
	\end{equation}
	which after integration yields,
	\begin{equation}  
	K_{Bulk} = 4\pi d^2 l_g \mathrm{e}^{b}+\int_{d}^{R_{Bj}} \mathrm{e}^{ \frac{\lambda_B}{r}}r^2 dr \,.
	\label{Eq18}
	\end{equation}
	The validity of the above equation is extended 
	beyond the strong coupling limit by replacing the integral with $K_{Eb}$. In the case of weak acids, large $l_g$,
	the first term will dominate Eq. (\ref{Eq18}), so that the bulk equilibrium constant for a weak acid can be approximated by 
	\begin{equation}
	K_{Bulk} = 4\pi d^2 l_g \mathrm{e}^{b},
	\label{ka}
	\end{equation}
	which is similar to Eq.~(\ref{eq:Keq2}) of the previous section.

	\section{Uniformly sticky colloid}
	\label{Ucol}
	To build a theory of charge regularization of colloidal particles we start with the simplest possible model in which the whole of colloidal surface is sticky.    Colloidal particle of radius $a$ is placed at the center of a spherical WS cell of radius $R$. The density profiles of ions,  then,  satisfy the modified Poisson-Boltzmann (mPB) equation: 
	%%%%%%%%%%%%%%%% equation %%%%%%%%%%%%%%%%%%%%%
	\begin{equation}  
	\nabla^2 \phi(r) =-\frac{4 \pi }{\epsilon_w}\sigma_0 \delta(r-a-r_{ion})  -\frac{4 \pi q}{\epsilon_w} \left[ c_{\ch{H^+}}(r) +c_+(r)- c_-(r) \right] ,
	\label{Eq12}
	\end{equation}   
	%%%%%%%%%%%%% end of equation %%%%%%%%%%%%%%%%%
	where $\sigma_0=0$ if the surface groups are basic, and $\sigma_0=-N_{sites} q/4 \pi (a+r_{ion})^2$ if all the groups are acidic.
	%%%%%%%%%%%%% end of equation %%%%%%%%%%%%%%%%%
	The ionic concentrations are defined as:
	%%%%%%%%%%%%%%%% equation %%%%%%%%%%%%%%%%%%%%%
	\begin{eqnarray}
	&&c_{\ch{H^+}}(r) = c_a~\mathrm{e}^{-\beta \left( u(r) +q \phi(r)\right)}\\ 
	&& c_+(r)= c_s~\mathrm{e}^{-\beta q \phi(r)}\\
	&&c_-(r) = (c_a+c_s)~\mathrm{e}^{ \beta q \phi(r)},	
	\label{Eq13}
	\label{Eq14}
	\label{Eq15}
	\end{eqnarray}  
	where $u(r)$ is the sticky potential between the colloidal surface and a hydronium ion,  $c_a=10^{-pH}$ is the reservoir concentration of  acid, and $c_s$ is the reservoir concentration of 1:1 salt.  We assume that both acid and salt in the reservoir are strong electrolytes and are fully dissociated.
	Using Eq.~\ref{eq:Bfactor} we obtain $\mathrm{e}^{-\beta \left( u(r) +q \phi(r)\right)}=(1-l_g \delta(r-a-r_{ion}))\mathrm{e}^{-\beta q \phi(r)}$,
	which means that the surface density of adsorbed hydronium ions is~\cite{bakhshandeh2019}:
	%%%%%%%%%%%%%%%% equation %%%%%%%%%%%%%%%%%%%%%
	\begin{equation}
	\begin{split}  
	\sigma_{su} =q c_a l_g \mathrm{e}^{-\beta q \phi_0},
	\end{split}
	\label{Eq16}
	\end{equation}  
	%%%%%%%%%%%%%%%%%%%%%%%%%%%%%%%%%%%%%%%%%%%%%%%%%%%%%%%%%%
	where $\phi_0=\phi(a+r_{ion})$. The net surface charge density is then
	\begin{equation}
	\sigma_{net}=  \sigma_0+  \sigma_{su}.
	\label{Eqa3}
	\end{equation}  
	To calculate the ionic density profiles and the number of condensed hydronium ions we must now solve the PB equation
	%%%%%%%%%%%%%%%% equation %%%%%%%%%%%%%%%%%%%%%
	\begin{equation}  
	\nabla^2 \phi(r) =  \frac{8 \pi q }{\epsilon_w} \left(c_a+c_s\right) \sinh[\beta \phi(r)] ,
	\label{Eq12a}
	\end{equation}   
	%%%%%%%%%%%%% end of equation %%%%%%%%%%%%%%%%%
	with the boundary conditions $\phi'(R)=0$ and $\phi'(a+r_{ion})=4\pi  \sigma_{net}/\epsilon_w$. 
	The calculation can be performed numerically using the 4th order Runge-Kutta,
	in which the value of the surface potential $\phi(a+r_{ion})=\phi_0$
	is adjusted based on the Newton-Raphson algorithm to obtain zero electric field at the cell boundary. 
	
	In realty, however,  the whole of colloidal surface is not uniformly sticky and hydronium ions can only adsorb  on special sites~\cite{jho2012}.  We now explicitly consider the modifications that must be made to the above theory in order to account for the discrete nature of adsorption sites.
	
	\section{Neutral functional groups}

	We first consider a colloidal particle with $N_{site}$ neutral basic groups (sticky spheres) uniformly distributed on its surface.
	To simplify the geometry we will map the spherical sticky sites onto  circular sticky
	patches of the same effective contact area.  Since both hydronium and the adsorption sites are modeled by spheres of the same diameter $d$, the hard core repulsion between the colloidal surface and the hydronium ion restricts the effective contact area to $2 \pi d^2$.  Therefore, the patch radius must be 
	%%%%%%%%%%%%%%%% equation %%%%%%%%%%%%%%%%%%%%%
	\begin{equation}
	\begin{split}  
	r_{patch} = \sqrt{2}~d,
	\end{split}
	\label{Eqa17}
	\end{equation}  
	%%%%%%%%%%%%%%%%%%%%%%%%%%%%%%%%%%%%%%%%%%%%%%%%%%%%%%%%%%
	Compared to the situation discussed in the previous section in which the whole of colloidal surface was sticky,
	the effective area on which hydronium ions can become adsorbed is significantly reduced in the case of discrete adsorption sites \cite{bakhshandeh2019}.  Nevertheless, we can still use the same approach as in Section \ref{Ucol}, if the sticky length is rescaled as $l_g^{eff}=l_g \alpha_{eff}$, to account for the reduced adsorption area, where
	\begin{equation}
	\alpha_{eff} = \frac{N_{site}^{act}\pi r_{patch}^2}{ 4 \pi (a+r_{ion})^2},  
	\label{Eqa18}
	\end{equation}  
	is the fraction of the surface area occupied by the {\it active} sticky patches. Note that
	if hydronium is adsorbed to a patch, this patch becomes inactive, preventing more than one hydronium ion from being adsorbed.  The number of adsorbed hydronium ion is given by Eq. (\ref{Eq16}) with $l_g$ replaced by $l_g^{eff}$.
	As the process of adsorption progresses, the number of active sites decreases in such a way as
	%%%%%%%%%%%%%%%% equation %%%%%%%%%%%%%%%%%%%%%
	\begin{equation}
	N_{site}^{act} =  N_{site}- 4\pi (a+r_{ion})^2 c_a l_g^{eff}~\mathrm{e}^{-\beta \phi_0},  
	\label{Eq19}
	\end{equation}
	resulting in a self-consistent equation for $l_g^{ eff}$ . 
	Solving Eqs. (\ref{Eqa18}) and (\ref{Eq19}), the effective sticky length is found to be
	%%%%%%%%%%%%%%%% equation %%%%%%%%%%%%%%%%%%%%%
	\begin{equation}  
	l_{g}^{eff} =\frac{l_g  N_{site} r_{patch}^2}{4  (a+r_{ion})^2\left(1+ l_g c_a \mathrm{e}^{-\beta \phi_0}\pi r_{patch}^2\right)}.
	\label{Eq20}
	\end{equation}
	The effective surface charge density 
	which must be used as the boundary condition for PB equation is then 
	%%%%%%%%%%%%%%%% equation %%%%%%%%%%%%%%%%%%%%%
	\begin{equation}  
	\sigma_{eff} =  q c_a l_g^{eff} \mathrm{e}^{-\beta \phi_0} = \frac{q K_{Surf} N_{site} c_a \mathrm{e}^{-\beta \phi_0}}{4 \pi (a+r_{ion})^2\left(1+ K_{Surf} c_a \mathrm{e}^{-\beta \phi_0}\right)}, 
	\label{Eq_neutral_1}
	\end{equation}   
	where $K_{Surf} =2 \pi l_g d^2= K_{Bulk}/2  $, where the bulk association constant is the same as in Eq.~\ref{Eq3-5}.
	
	We stress again that the bulk  equilibrium  constant $K_{Bulk}$ is exact for the model of sticky hard spheres and does not depend on the density of the reactants.  The higher order terms of the virial expansion, however, will modify the activity coefficients, so that 
	in the law of mass action the concentrations will have to be replaced by the activities. Nevertheless, since the PB equation does not take into account ionic correlations,
	to remain consistent, the activity coefficients must also be set to unity. Solving Eq.~\ref{Eq12a}, with the boundary conditions $\phi'(R)=0$ and $\phi'(a+r_{ion})=4\pi  \sigma_{net}/\epsilon_w$ we are able to obtain the density profile of ions around the colloidal particle. 
	
\begin{figure}
		\begin{center}
			\includegraphics[width=7cm]{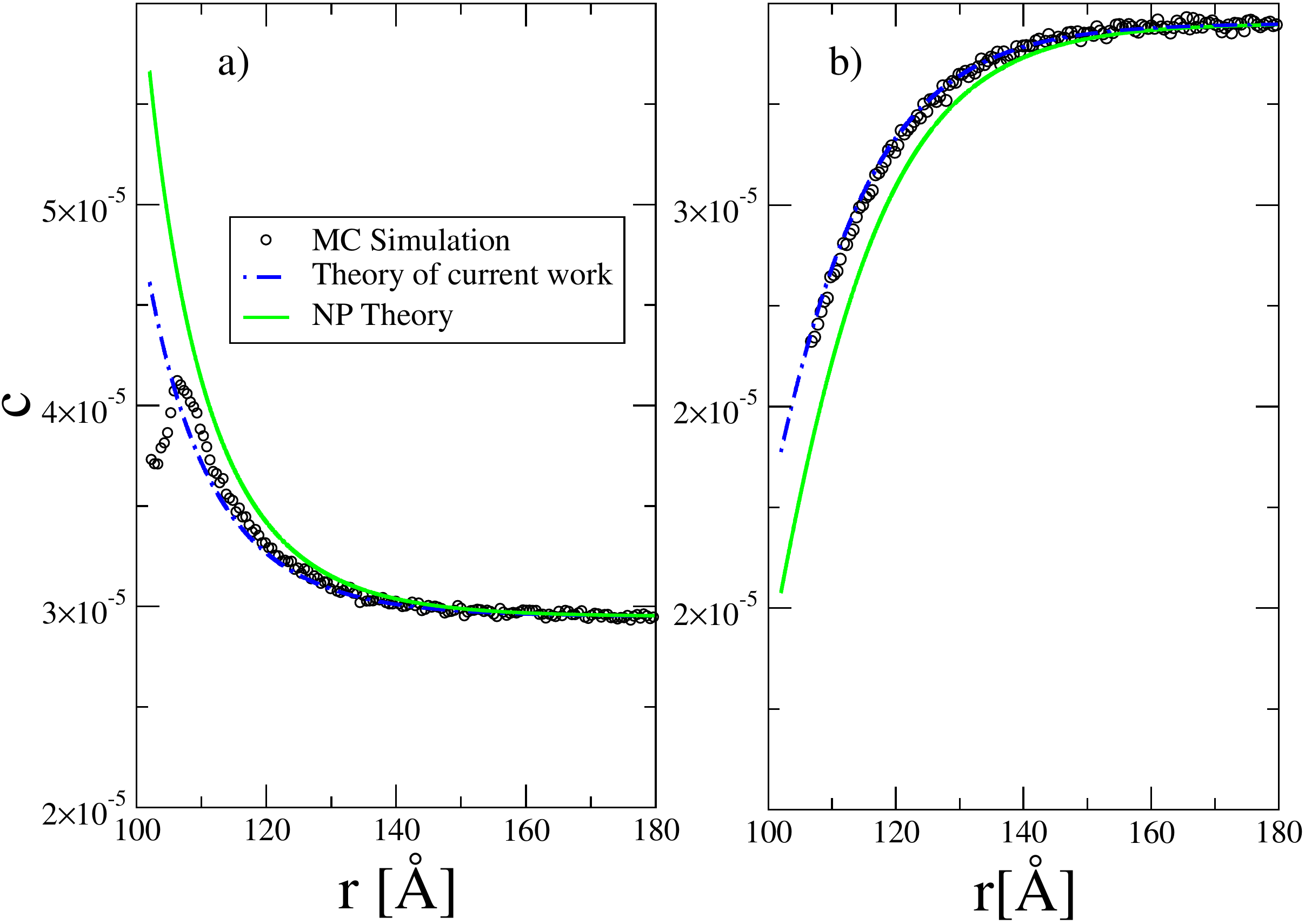}
		\end{center}
		\caption{Density profiles of hydronium and \ch{Cl-} measured in
			particles per \AA$^3$ . Symbols are the simulation data and solid (green)
			and dashed (blue) lines are the predictions of the NP theory and of the theory developed in the present work, respectively. The parameters are $a =100$ \AA, $R = 200$ \AA, and $l_g=109.97 $\AA. The colloidal particle has $300$ neutral basic sites
			on its surface. The concentration of \ch{HCl} is $50$ mM. a)Density profile of  \ch{Cl-} and b)  Density profile of hydronium.}  
		\label{fig2}
	\end{figure}

\begin{figure}
		\begin{center}
			\includegraphics[width=7cm]{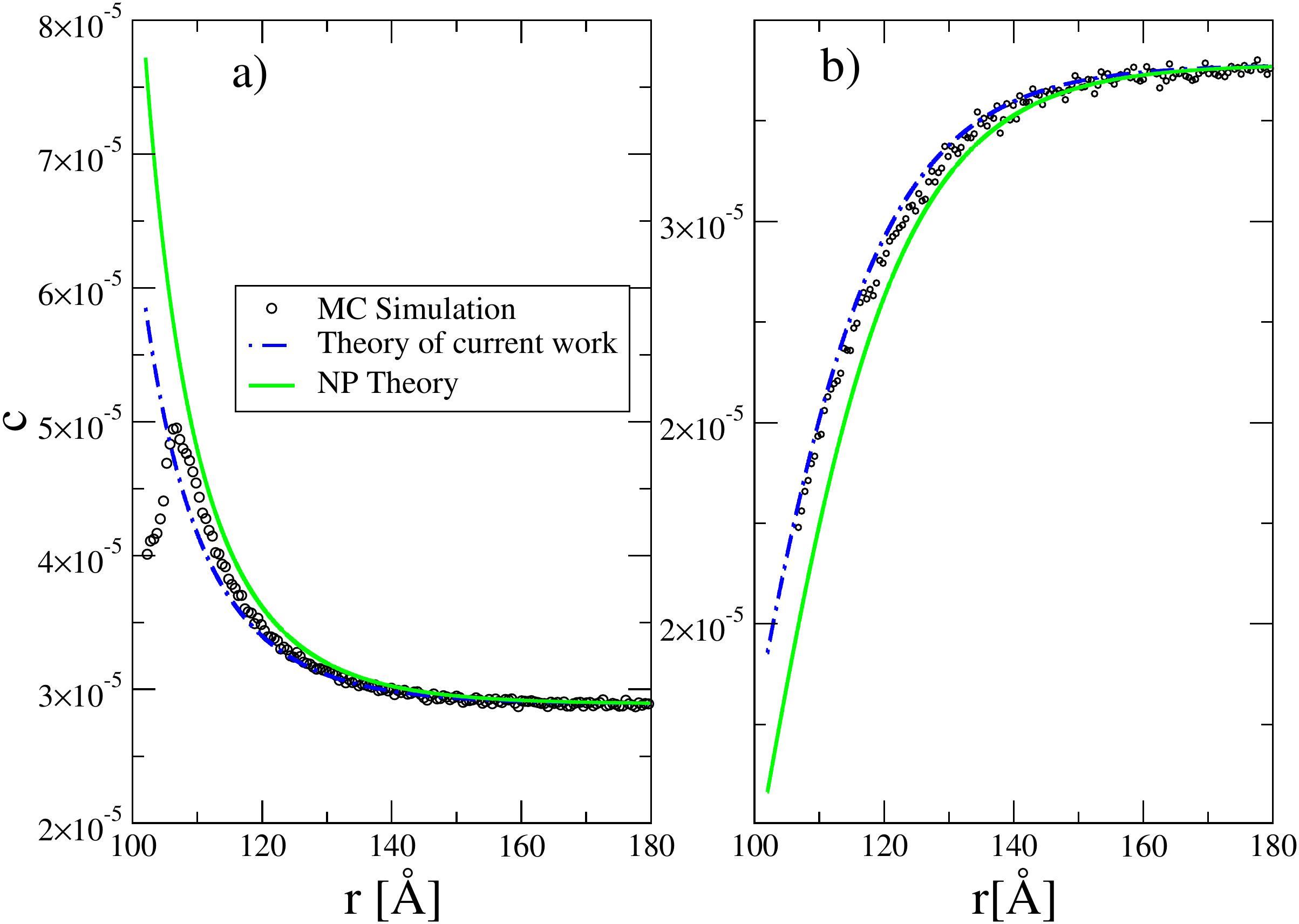}
		\end{center}
		\caption{Density profiles of hydronium and \ch{Cl-} measured in
			particles per \AA$^3$ . Symbols are the simulation data and solid (green)
			and dashed (blue) lines are the predictions of the NP theory and of the theory developed in the present work, respectively. The parameters are $a =100$ \AA, $R = 200$ \AA, and $l_g=109.97 $\AA. The colloidal particle has $600$ basic sites
			on its surface. The concentration of \ch{HCl} is $50$ mM. a)Density profile of  \ch{Cl-} and b)  Density profile of hydronium.   }  
		\label{fig3}
	\end{figure}

	To explore the range of validity of the theory we will compare it with the results of Monte Carlo simulations.
	We first consider colloidal particles with $300$ and $600$ active neutral basic sites and concentration of HCl set to $50$ mM. In Figs.~\ref{fig2} and ~\ref{fig3} we show the comparison between the simulation data, NP theory, and the present work,
	For these parameters the difference between the new theory and the NP approach is not very large, nevertheless it is clear that the simulation results are in a much better agreement with the theory developed in the present paper. The figures show that the density of free hydronium ions decreases near the colloidal surface.  This is not surprising, since once some of the hydroniums have adsorbed to the neutral basic groups, colloidal surface becomes positively charged and repels other cations.  A more curious behavior is found for the anion \ch{Cl-}, the concentration of which shows a peak close to the surface, but then diminishes on further approach.    
	The reason for this  is that anions prefer to stay close to the adsorbed  cations \ch{H+}, which in turn want to minimize the repulsive electrostatic energy between themselves, as well as to maximize entropy.  This favors the hydronium ions to be located at about $~3~r_{ion}$ from the colloidal surface.  This is precisely the position of the peak found in the density profile of anions. This fine detail, however, is beyond the scope of the present theory.  Nevertheless the fact that the density profiles away from colloidal surface are perfectly described by the present theory implies that the prediction for the total number of adsorbed hydronium ions is correct, in spite of the fine structure of ionic density profiles near the surface.  
	%%%%%%%%%%%%%%%%% figure %%%%%%%%%%%%%%%%%%%%%
	%%%%%%%%%%%%% end of figure %%%%%%%%%%%%%%%%%
	
	%%%%%%%%%%%%%%%%% figure %%%%%%%%%%%%%%%%%%%%%
	%%%%%%%%%%%%% end of figure %%%%%%%%%%%%%%%%%

\begin{figure}
		\begin{center}
			\includegraphics[width=7cm]{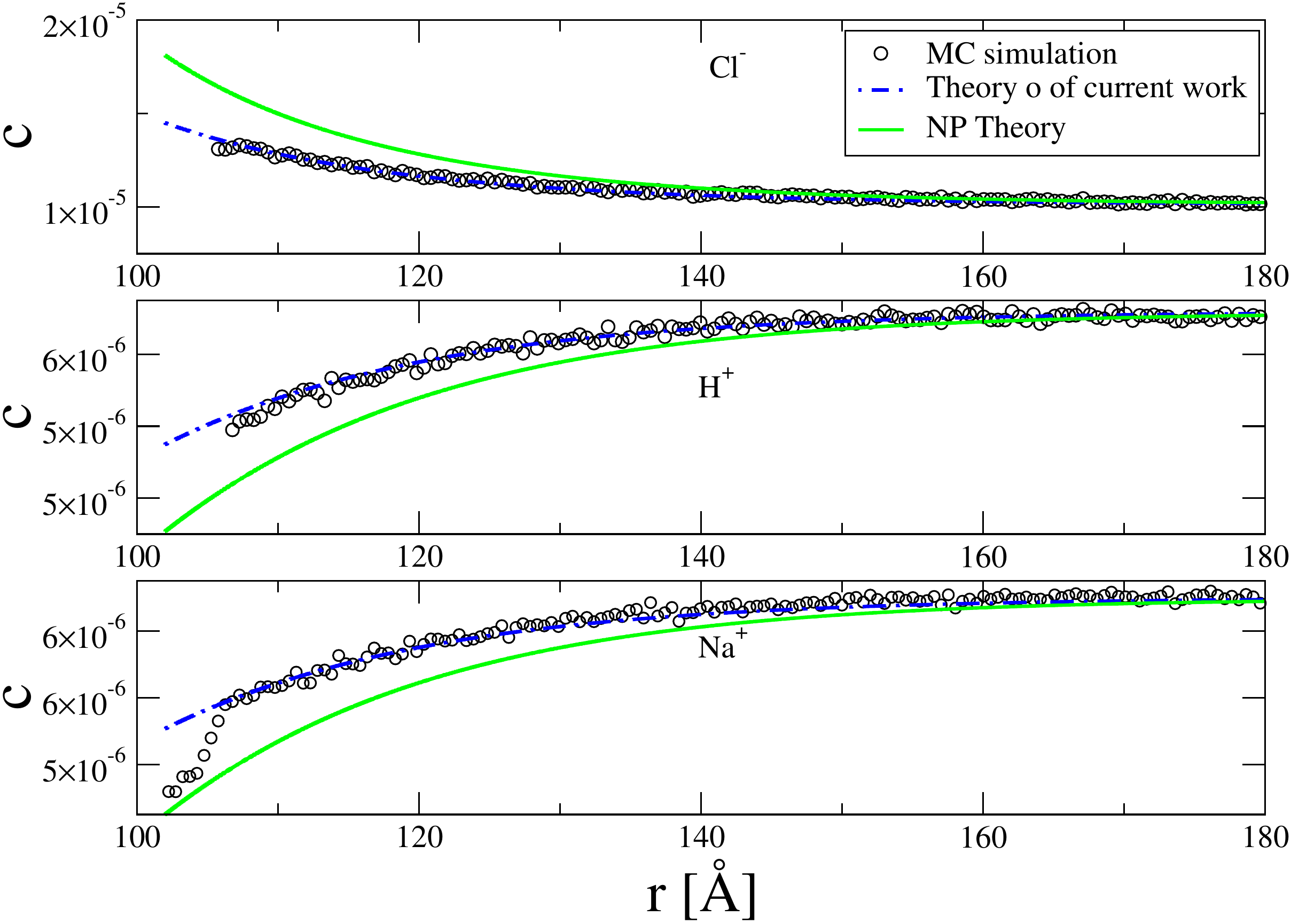}
		\end{center}
		\caption{Density profiles of hydronium, \ch{Cl-}, and \ch{Na+} measured in
			particles per \AA$^3$ . Symbols are the simulation data and solid (green)
			and dashed (blue) lines are the predictions of the NP theory and of the theory developed in the present work, respectively. The parameters are $a =100$ \AA, $R = 200$ \AA, and $l_g=109.97 $\AA. The colloidal particle has $200$ basic sites
			on its surface.  The concentration of \ch{HCl} and \ch{NaCl}  is $10$ mM. The density C is in units of
			particles per \AA$^3$.}  
		\label{fig4}
	\end{figure}

	We next consider the effect of 1:1 salt on the charge regulation.  We study a colloidal particle with $200$ basic sites in the presence of \ch{HCl} and \ch{NaCl}, both at concentration $10$ mM.  We assume that both acid and salt are completely ionized. The results of the theory and simulations are shown in Fig.~\ref{fig4}.
	%%%%%%%%%%%%%%%%% figure %%%%%%%%%%%%%%%%%%%%%
	%%%%%%%%%%%%% end of figure %%%%%%%%%%%%%%%%%
	Once again we see a very good agreement between the present theory and the MCs simulations. 
	
\begin{figure}
		\begin{center}
			\includegraphics[width=7cm]{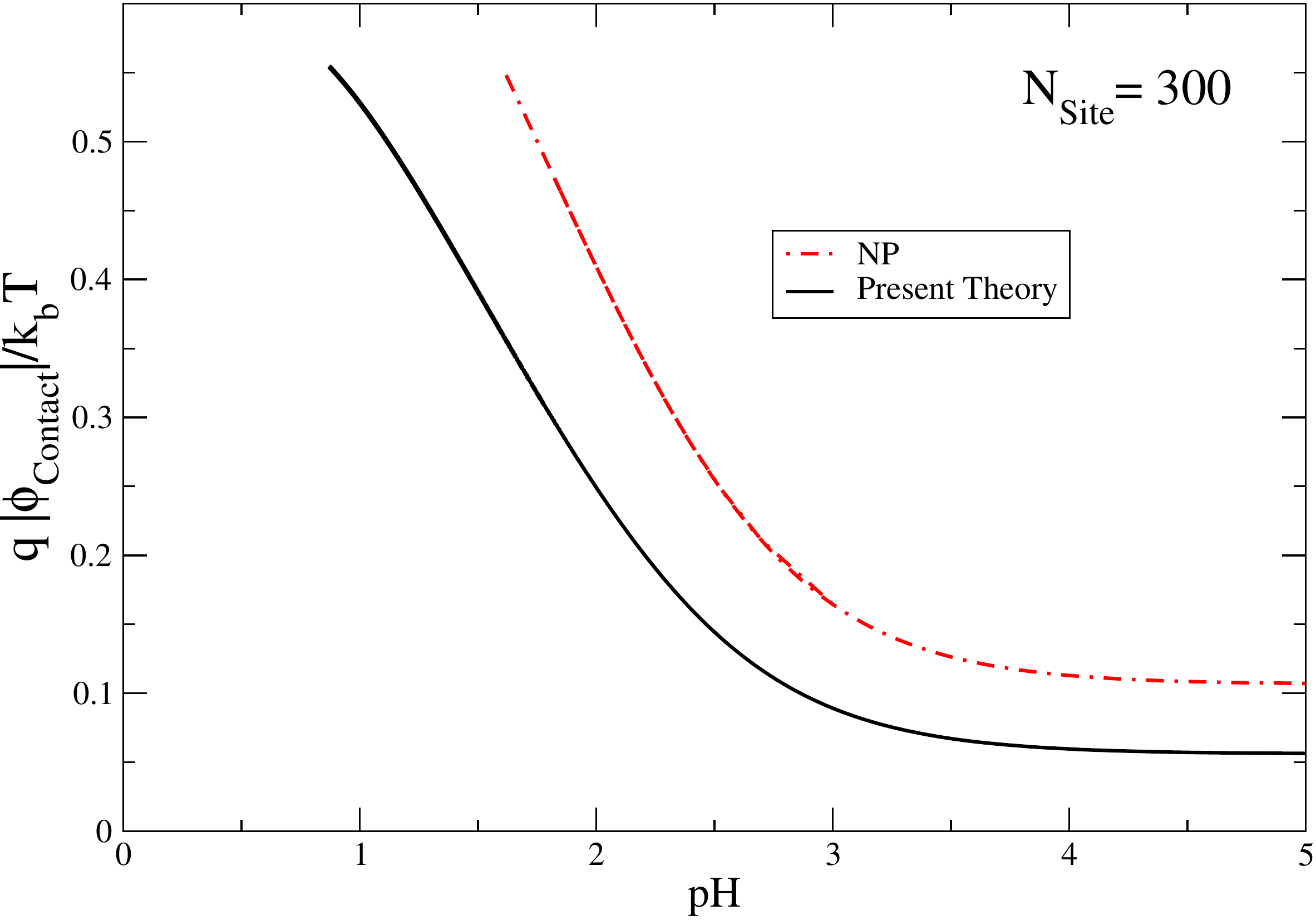}
		\end{center}
		\caption{Contact potential as a function of pH for the present theory and NP theory, respectively. The colloidal particle has $300$ basic functional
			group on it surface. The parameters are $a =100$ \AA, $R = 200$ \AA, and $l_g=109.97 $\AA.  There is no added salt. 
		}  
		\label{fig5}
	\end{figure}

\begin{figure}
		\begin{center}
			\includegraphics[width=7cm]{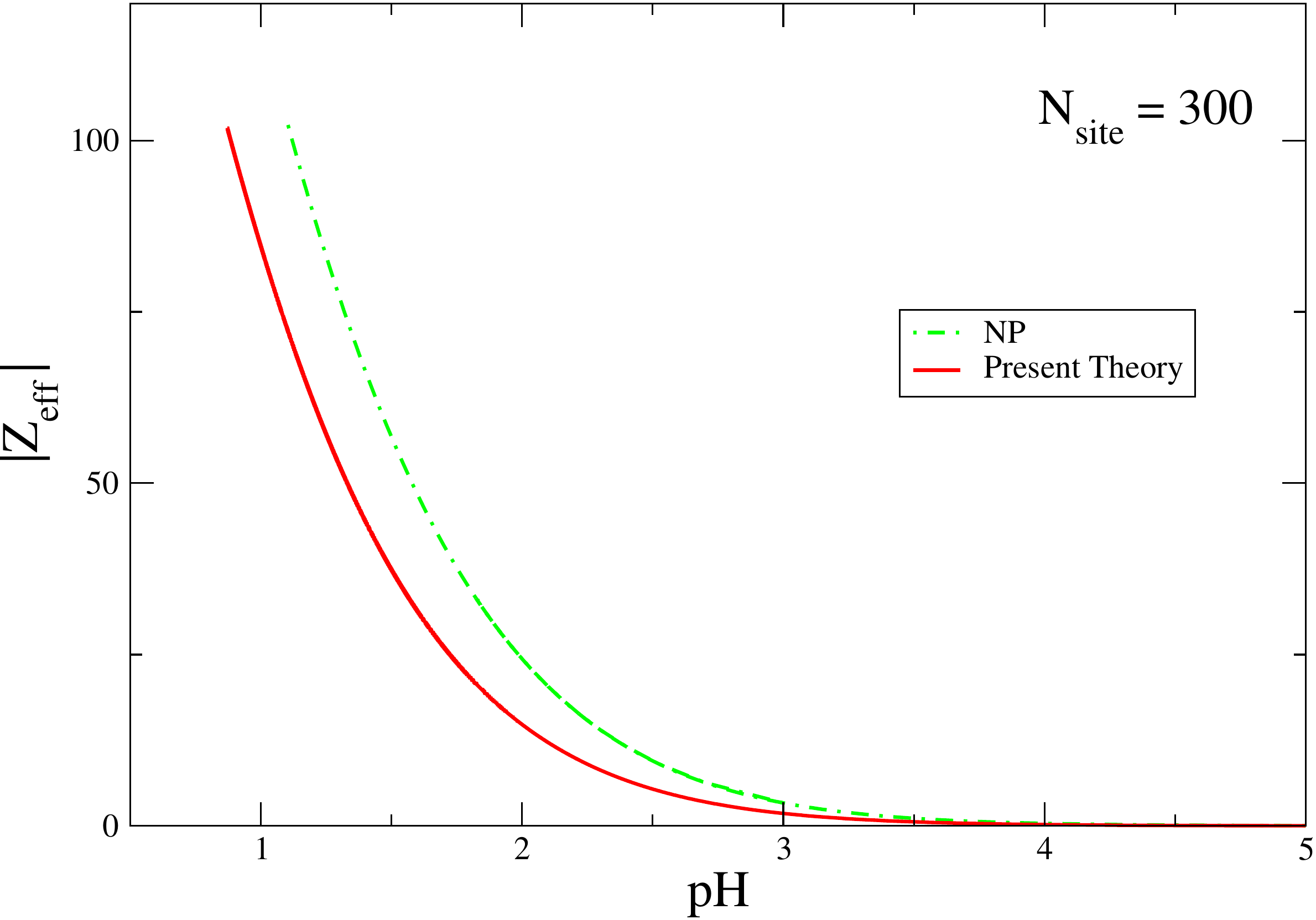}
		\end{center}
		\caption{Effective charge of colloidal particle in unit of $q$ as a function of pH for the present theory and NP theory, respectively. The colloidal particle has $300$ basic active functional
			group on it surface. The parameters are $a =100$ \AA, $R = 200$ \AA, and $l_g=109.97 $\AA. There is no added salt.
		}  
		\label{fig6}
	\end{figure}

\begin{figure}
		%%%%%%%%%%%%%%%%%%%%%%%%%%%%%%%%%%%%%%%%%%%%%%%%	
		\begin{center}
			\includegraphics[width=7cm]{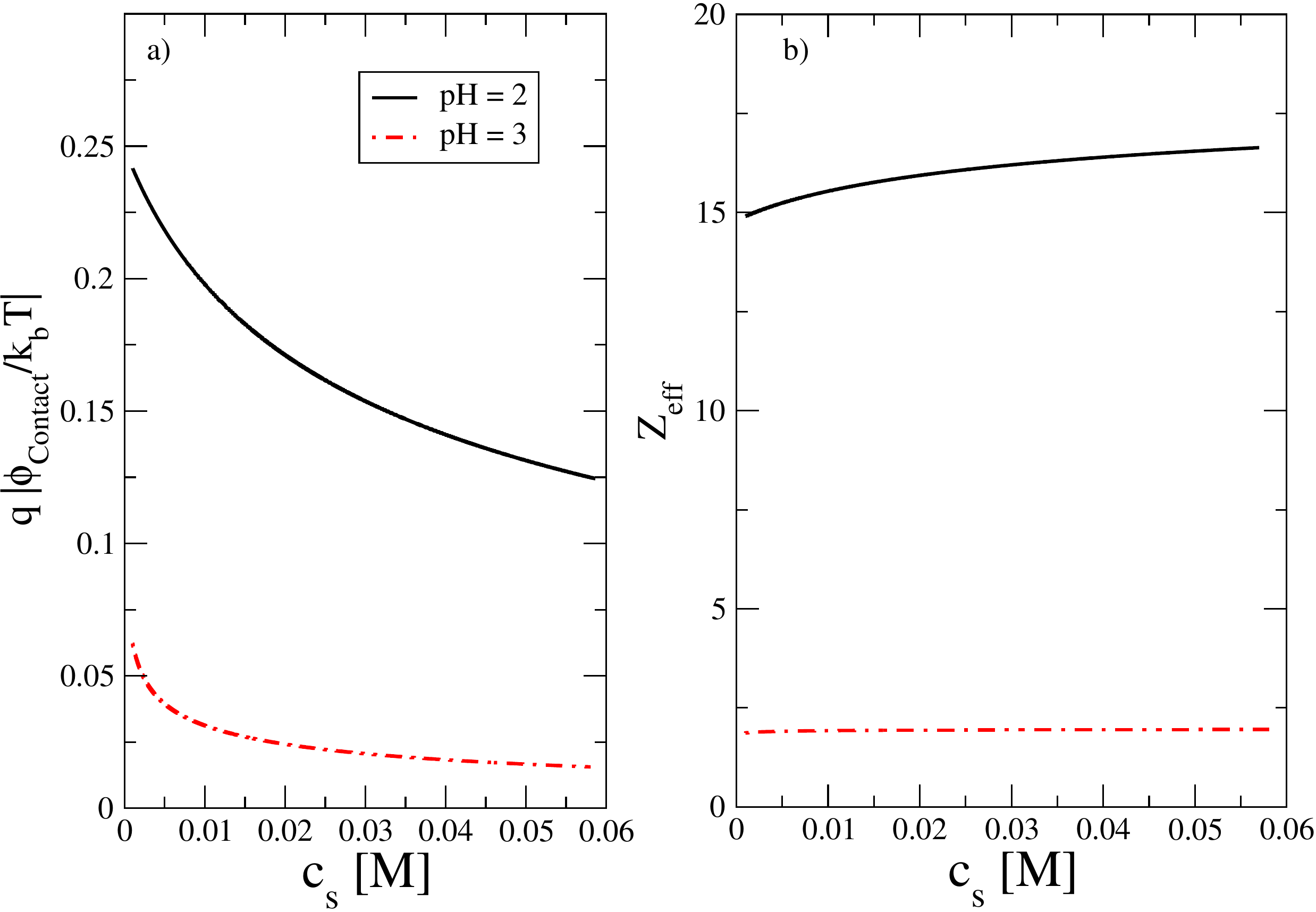}
		\end{center}
		\caption{(a) Contact potential  and (b) the effect charge  of colloidal particle in unit of $q$ as a function of salt concentration  $c_s$ (M) for different  pH. The colloidal particle has $300$ basic functional
			group. The parameters are $a =100$ \AA, $R = 200$ \AA, and $l_g=109.97 $\AA.  
		}  
		\label{fadd}
	\end{figure}

	In experiments, Zeta potential is more easily available than the effective charge. Definition of Zeta potential, however, requires knowledge of the position of the slip plain. Nevertheless we expect that Zeta potential  will behave similarly to the  electrostatic contact surface potential. 
	In Figs.~\ref{fig5} and \ref{fig6} we show the behavior of the surface potential and the effective charge $Z_{eff}$ in unit of charge $q$  as a function of pH,  for the present theory and NP theory, respectively
	and in Fig~\ref{fadd}, the behavior of the two as a function of salt concentration. We observe that addition of 1:1 electrolyte diminishes the contact potential. This, in turn, lowers the electrostatic  energy penalty for bringing hydronium ions to colloidal surface, thus favoring their association with the active sites.  Indeed, Fig~\ref{fadd}b shows that the effective charge of colloidal particle increases with increasing salt concentration. 
	%%%%%%%%%%%%%%%%% figure %%%%%%%%%%%%%%%%%%%%%
	%%%%%%%%%%%%% end of figure %%%%%%%%%%%%%%%%%
	%Fig.~\ref{fig5} shows that the contact potential increases with increasing pH as expected, this is in agreement with Fig.~\ref{fig6} which shows the increasing of effective charge of colloidal particle.

	\section{Charged functional groups}
	
	We next consider a colloidal particle with $N_{site}$ acidic surface groups each carrying a charge $-q$.  If all the groups would be ionized, the particle would acquire a net charge $Q_0=-N_{site} q$.  The chemical equilibrium between hydronium and acid groups, however, reduces this value to $Q_{eff}=Q_0+Q_{con}$, where 
	\begin{equation}
	Q_{con}=4  \pi (a+r_{ion})^2 q c_a l_g^{eff} \mathrm{e}^{-\beta q \varphi_0}
	\end{equation}
	is the number of associated hydronium ion and $\varphi_0$ is the potential of mean force (PMF) --- the work required to bring an ion from the bulk to contact with
	one of the acidic groups. The PMF can be separated into a mean-field electrostatic potential $\phi_0$
	and a contribution from the discrete nature of surface charge groups $\mu^{qq}_c$,
	\begin{equation}  
	\varphi_0 = \phi_0 +\mu^{qq}_c \,.
	\label{Eqadd1}
	\end{equation}
	The value of $l_g^{eff}$ is given by Eq.~(\ref{Eq20}) with the mean-field potential replaced by the PMF, $\phi_0 \rightarrow \varphi_0$.
	The effective surface charge density then reduces to 
	\begin{equation}  
	\sigma_{eff} = -\frac{N_{site} q}{4 \pi\left(a+r_{ion}\right)^2} +\frac{q K_{Surf}^a  N_{site}c_a \mathrm{e}^{-\beta \phi_0}}{4 \pi (a+r_{ion})^2\left(1+ K_{Surf}^a c_a \mathrm{e}^{-\beta \phi_0}\right)} 
	\label{Eqadd3}
	\end{equation}
	where 
	\begin{equation}
	K_{Surf}^a = \frac{K_{Bulk}^a}{2}\mathrm{e}^{-b-\beta \mu^{qq}_c},
	\label{ksur}
	\end{equation}
	and the bulk acid association constant $K_{Bulk}^a$ is given by Eq.~(\ref{ka}). The term $\mathrm{e}^{-b}$ in Eq.~\ref{ksur}  discounts the direct Coulomb interaction between the hydronium ion and its adsorption site, which is already accounted for in the $ \mu^{qq}_c$.

	\section{The effect of discrete charges}
	It is well known that the PB equation is very accurate for systems containing only 1:1 electrolyte.
	The mean-field nature of this equation is manifested by the complete neglect of ionic correlations, 
	which are found to be small for aqueous solutions of monovalent ions~\cite{levin}.  
	However, in the case of acidic groups, hydronium ions will condense directly onto charged sites
	and discrete nature of hydronium ions and surface sites can not be neglected for the {\it associated} ions.
	The free ions, however, can still be treated at the mean-field level.  
	
	To account for the discrete nature of surface groups, we add and subtract a uniform neutralizing background to the colloidal surface.  The negative of the background can be combined with 
	the mean-field electrostatic potential produced by the ions
	to yield the total mean-field electrostatic potential $\phi(r)$. The potential produced by the discrete surface charge and their neutralizing background, on the other hand, correspond to $\mu^{qq}_c$ defined in Eq.(\ref{Eqadd1}).  To calculate $\mu^{qq}_c$ we will ignore the curvature of the colloidal surface. Furthermore, we will suppose that the adsorption sites are uniformly distributed, forming a triangular lattice of spacing $L$.    
	
\begin{figure}
		\begin{center}
			\includegraphics[width=7cm]{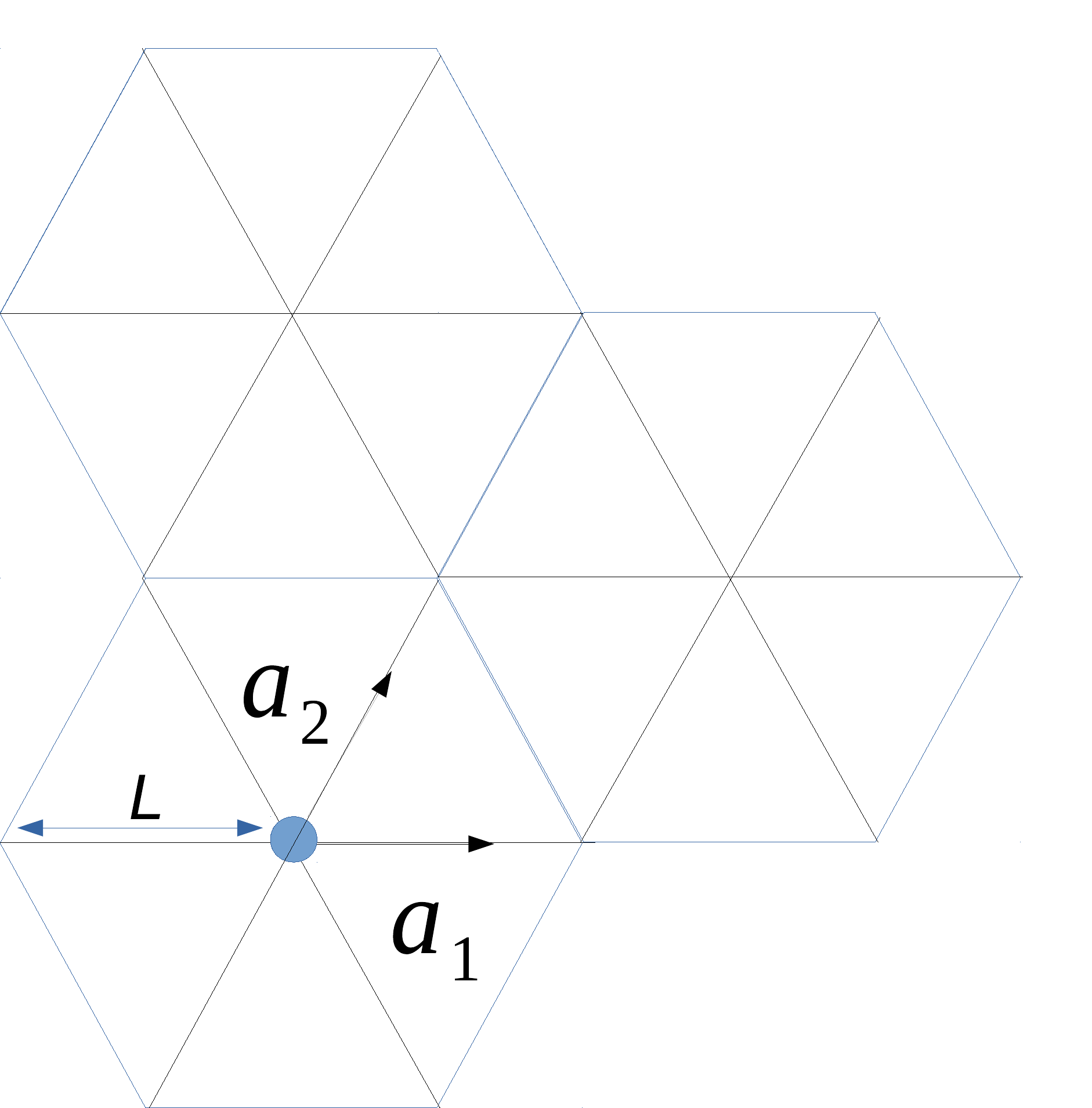}
		\end{center}
		\caption{The triangular lattice used to evaluate $\mu^{qq}_{c}$ and $\mu^{qq}_{n}$. 
		}  
		\label{lattice}
	\end{figure}

	We start by calculating the electrostatic potential produced by an infinite planar triangular array of charges, see Fig.~\ref{lattice}.  This potential must satisfy the Poisson equation
	\begin{equation}
	\nabla^2 G(\boldsymbol{r}) = -\frac{4 \pi q}{\epsilon_w}  \sum_{n,m} \delta(z)\delta(\boldsymbol{\rho} -n \boldsymbol{a}_1-m \boldsymbol{a}_2),
	\label{pois}
	\end{equation}
	%%%%%%%%%%%%% end of figure %%%%%%%%%%%%%%%%%
	%%%%%%%%%%%%%%%%%%%%%%%%%%%
	where $z$ and $\boldsymbol{\rho} = x \boldsymbol{\hat{x}} + y \boldsymbol{\hat{y}}$ are the transverse and longitudinal directions, respectively, and the lattice vectors are given by
	\begin{equation}
	\begin{split}
	\boldsymbol{a}_1 = L~\boldsymbol{\hat{x}}  ,\\
	\boldsymbol{a}_2 = \frac{1}{2} L~ \boldsymbol{\hat{x}} + \frac{\sqrt{3}}{2} L\boldsymbol{\hat{y}}. 
	\end{split}
	\label{vector}
	\end{equation}
	
	The area of the unit cell  of triangular lattice is
	\begin{equation}
	\begin{split}
	\abs{\gamma } =	\abs{\boldsymbol{a}_1 \times 	\boldsymbol{a}_2}=  \frac{\sqrt{3}}{2} L^2
	\end{split}
	\label{v1}
	\end{equation}
	The reciprocal lattice vectors $\boldsymbol{b}_i$ are defined as $\boldsymbol{a}_j \cdot \boldsymbol{b}_j = 2\pi \delta_{ij}$,
	and are given by
	\begin{equation}
	\begin{split}
	\boldsymbol{b}_1 = \frac{2 \pi}{d} \left(\boldsymbol{\hat{x}} - \frac{\boldsymbol{\hat{y}}}{\sqrt{3}}\right)	.\\ 	
	\boldsymbol{b}_2 = \frac{2 \pi}{d} \left( \frac{2 \boldsymbol{\hat{y}}}{\sqrt{3}}\right). 
	\end{split}
	\label{v2}
	\end{equation}
	The periodic delta function can be written as 
	\begin{equation}
	\begin{split}
	\sum_{n,m} \delta(\boldsymbol{\rho} -n \boldsymbol{a}_1-m \boldsymbol{a}_2)=\\	 	
	\frac{1}{\gamma}  \sum_{n,m} \mathrm{e}^{i\boldsymbol{b}_1\cdot \boldsymbol{\rho}n +i \boldsymbol{b}_2\cdot \boldsymbol{\rho}m }
	\end{split}
	\label{v3}
	\end{equation}
	and the Green function as~\cite{SaGi17}
	\begin{equation}
	\begin{split}
	G(\boldsymbol{r}) = \frac{1}{\gamma}  \sum_{n,m} g_{n,m}(z)  \mathrm{e}^{i \boldsymbol{b}_1\cdot \boldsymbol{\rho}n +i \boldsymbol{b}_2\cdot \boldsymbol{\rho}m },
	\end{split}
	\label{v4}
	\end{equation}
	where $g_{n,m}(z) $ is a function of $z$ coordinate only.
	Substituting Eq. (\ref{v4}) into Eq. (\ref{pois}) we obtain
	\begin{equation}
	\begin{split}
	\frac{\partial^2 g_{n,m}(z)}{\partial z^2} -k^2 g_{n,m}(z)=-\frac{4 \pi q}{\epsilon_w}\delta(z),\\
	k = \sqrt{\frac{4 \pi^2}{L^2}\left(n^2 +\left(\frac{2 m }{\sqrt{3}} -\frac{n}{\sqrt{3}}\right)^2\right)},
	\end{split}
	\label{v6}
	\end{equation}
	which has a solution of the form
	\begin{equation}  
	g_{n,m}(z) =
	\begin{cases}
	A~\mathrm{e}^{-k z}, & z>0,\\
	A~\mathrm{e}^{k z}, &       z<0,
	\end{cases} 
	\label{v7}  
	\end{equation}  
	Integrating Eq.~(\ref{v6}) once, we see that the derivative of $g(z)$ is discontinuous at $z=0$ with
	\begin{equation}  
	g^{\prime}_{n,m}(0^+)-g^{\prime}_{n,m}(0^-) =-\frac{4 \pi q}{\epsilon_w},     
	\label{v8}  
	\end{equation} 
	from which we determine $A=2 \pi q/\epsilon_w k$,. The Green function can then be written as
	\begin{equation}
	\begin{split}
	G(\boldsymbol{r}) = \frac{2 \pi q}{\gamma \epsilon_w}  \sum_{n=-\infty}^{n=\infty} \sum_{m=-\infty}^{m=\infty} \\
	\frac{\mathrm{e}^{-k\abs{z}}}{k}\cos{\frac{2 \pi}{L}\left(n x+ \frac{1}{\sqrt{3}}(2y m - y n) \right)}.
	\end{split}
	\label{v10}
	\end{equation}
	The $(n=0,m=0)$ term of  $ G(\boldsymbol{r})$ diverges. Indeed, if we take the limit $k \rightarrow 0$ of the summation and 
	in Eq. (\ref{v10}) we will obtain
	an infinite constant and a finite term which grows as $\abs{z}$.  This is nothing more than the potential of a uniformly charged plane. Therefore, if we introduce a neutralizing background, we will cancel precisely this term, eliminating the divergence.  The electrostatic potential produced by a triangular array of charges {\it on a neutralizing background} is then
	\begin{equation}
	\begin{split}
	\bar G(\boldsymbol{r}) = \frac{2 \pi q}{\gamma \epsilon_w}  \sum_{n=-\infty}^{n=\infty \prime} \sum_{m=-\infty}^{m=\infty \prime} 
	\frac{\mathrm{e}^{-k\abs{z}}}{k}\\
	\cos{\frac{2 \pi}{L}\left(n x+ \frac{1}{\sqrt{3}}(2y m - y n) \right)},
	\end{split}
	\label{v10a}
	\end{equation}
	where the prime on the sums indicates that we have removed the term $(n=0,m=0)$. 
	Bringing an ion of opposite charge into contact with one of the adsorption sites then yield    
	\begin{equation}
	\begin{split}
	\mu^{qq}_c = -\frac{2 \pi q^2}{\gamma \epsilon_w}\sum_{n=-\infty}^{n=\infty \prime} \sum_{m=-\infty}^{m=\infty \prime} \frac{\mathrm{e}^{-2 k r_{ion}}}{k}.
	\end{split}
	\label{v11}
	\end{equation}
	Even if sites are not perfectly ordered on the colloidal surface, we still expect that $\mu^{qq}_c$
	derived in Eq. (\ref{v11}) will provide a reasonably accurate account of the discreteness effects assuming that
	the average separation between $Z$ acid sites is such that the area per site is $\gamma=4 \pi (a+r_{ion})^2/Z$, where $\gamma$ is given by Eq.~\ref{v1}. The average separation between acid groups is then $L=(a+r_{ion}) \sqrt{8 \pi/\sqrt{3} Z}$
	
	%At this point, we are able to evaluate density profile of ions around colloidal surface using Eq.~\ref{Eqadd3}. 
	%In our model, as soon as a hydronium ion adsorbs to the site, the site becomes inactive and other hydronium cannot  react and adsorb to that site and if the site is charged only interaction with other ions is electrostatic.
	
\begin{figure}
		\begin{center}
			\includegraphics[width=7cm]{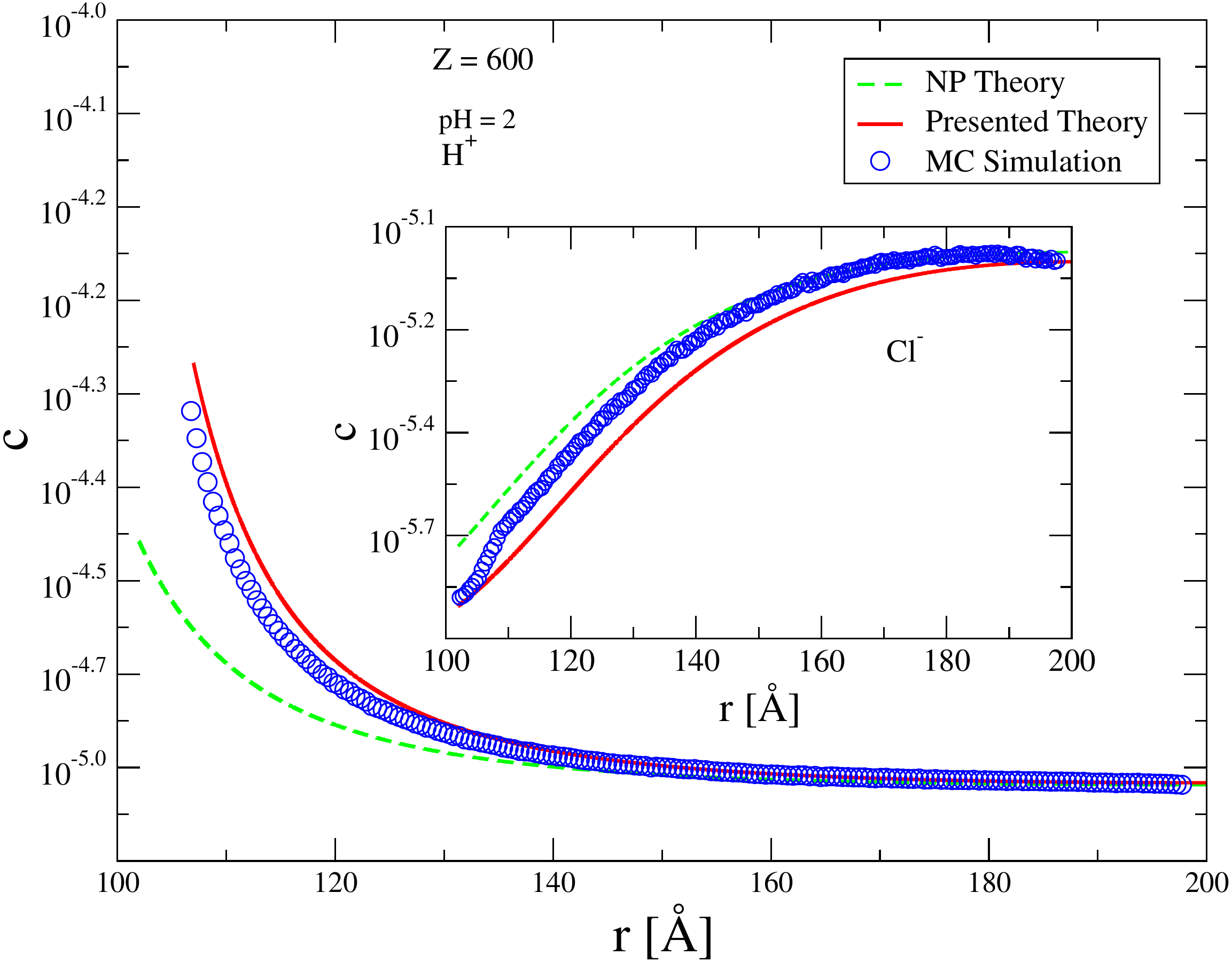}
		\end{center}
		\caption{Comparison between the present theory (solid lines), NP theory (dashed lines) and simulations
			(symbols), for colloidal particles with $Z=600$ and $l_g=109.97 $\AA$\,$ functional groups. The densities are in units of
			particles per \AA$^3$
		}  
		\label{fig_charged_1}
	\end{figure}

\begin{figure}
		\begin{center}
			\includegraphics[width=7cm]{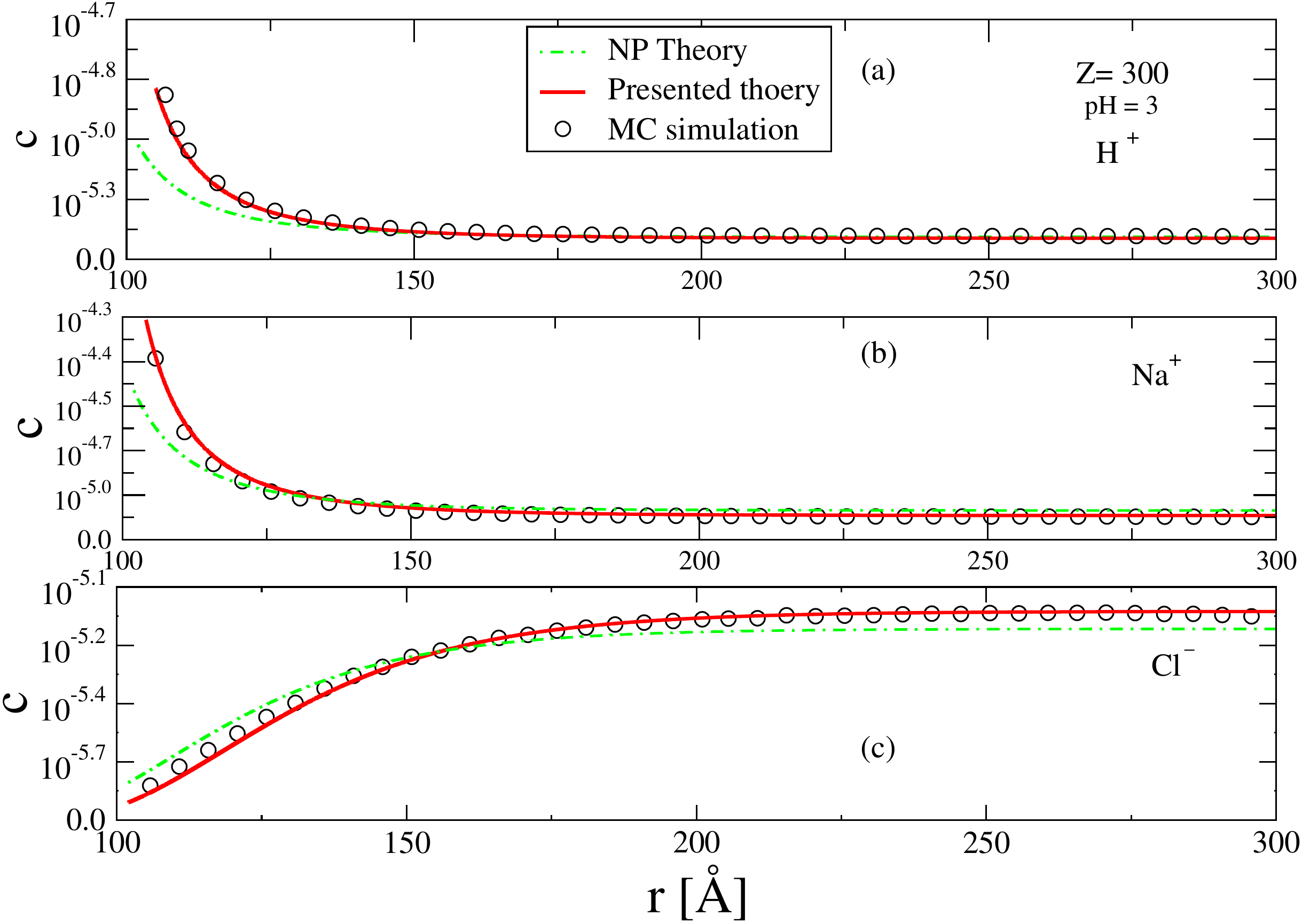}
		\end{center}
		\caption{
			Comparison between the present theory (solid lines), NP theory (dashed lines) and simulations
			(symbols), for colloidal particles with $Z=300$ charged functional group with $l_g=109.97 $\AA.  Solution is at pH $=3$ and has $10$mM bulk $1:1$ salt concentration. The densities are in units of
			particles per \AA$^3$.
		}  
		\label{fig_charged_2}
	\end{figure}

	We first consider a colloidal particles with $600$ acid surface groups with $l_g=109.97$ \AA, in a solution of pH = $2$. The ionic density profiles are presented in Fig.~\ref{fig_charged_1}. We see that the theory is in excellent agreement with simulations, while NP approach shows significant deviation. Next we consider particles with $300$ charged sites inside an acid solution containing 1:1 salt. Once gain there is a good agreement between theory and simulations, see  Fig.~\ref{fig_charged_2}.
	
	%%%%%%%%%%
	%%%%%%%%%%
	
\begin{figure}
		\begin{center}
			\includegraphics[width=7cm]{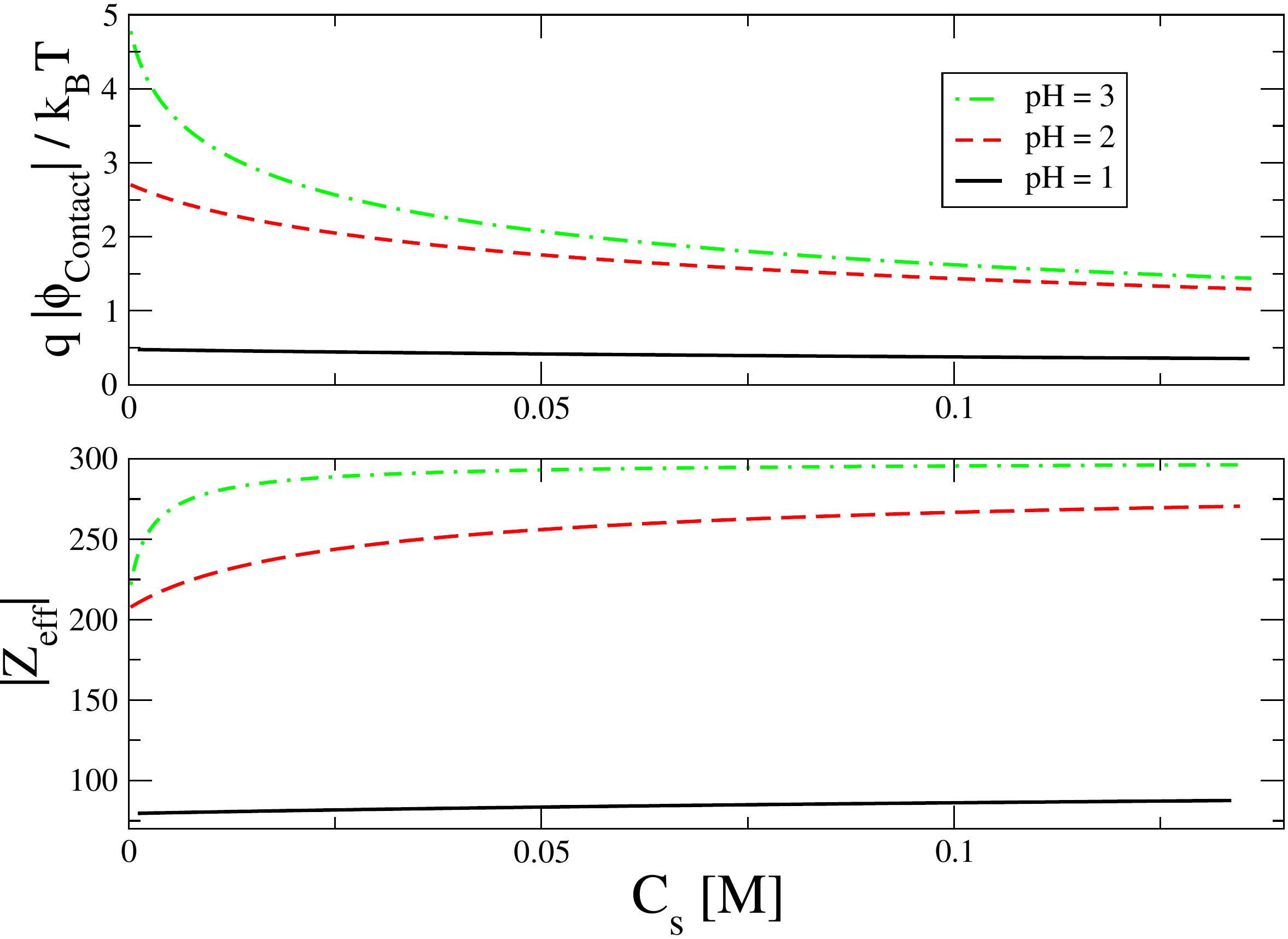}
		\end{center}
		\caption{Modulus of the effective charge in unit of $q$ and contact potential of colloidal particle as a function of 1:1 salt concentration ~ $C_s$    for different values of pH. The colloidal particle has $300$ charged  functional group on it surface. The parameters are $a =100$ \AA, $R = 200$ \AA, and $l_g=109.97 $\AA. The densities are in units of
			particles per \AA$^3$ 
		}  
		\label{fig7}
	\end{figure}

\begin{figure}
		\begin{center}
			\includegraphics[width=7cm]{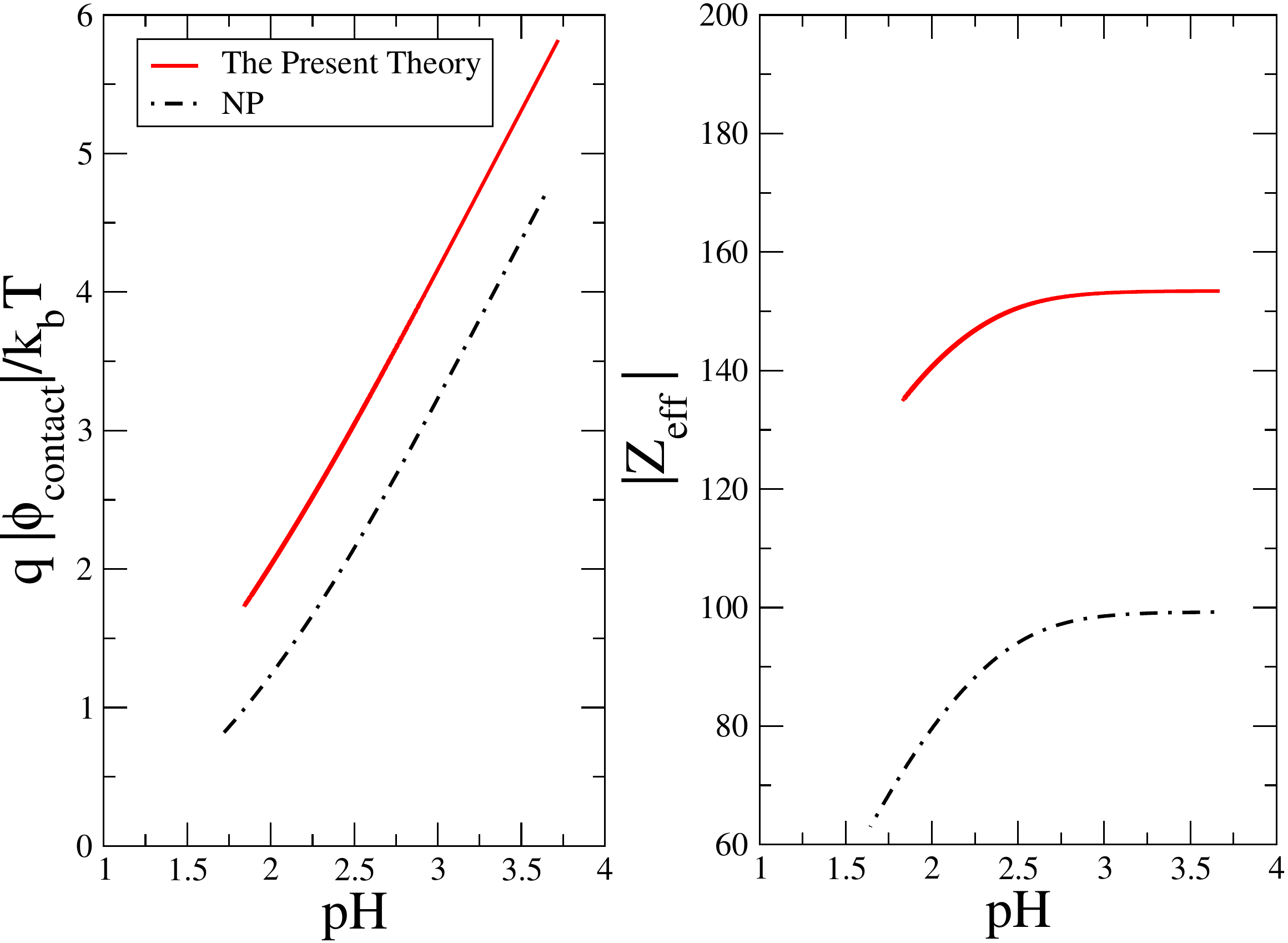}
		\end{center}
		\caption{{The modulus of the effective charge in units of $q$ and the contact potential of a nanoparticle as a function of pH in the acid reservoir, predicted by the NP and the present theories. The colloidal particle has $300$ charged  functional group on it surface. The parameters are $a =100$ \AA, $R = 200$ \AA, and $l_g=109.97 $\AA.  The system is salt-free. } 
		}  
		\label{figadded}
	\end{figure}

\begin{figure}
		\begin{center}
			\includegraphics[width=7cm]{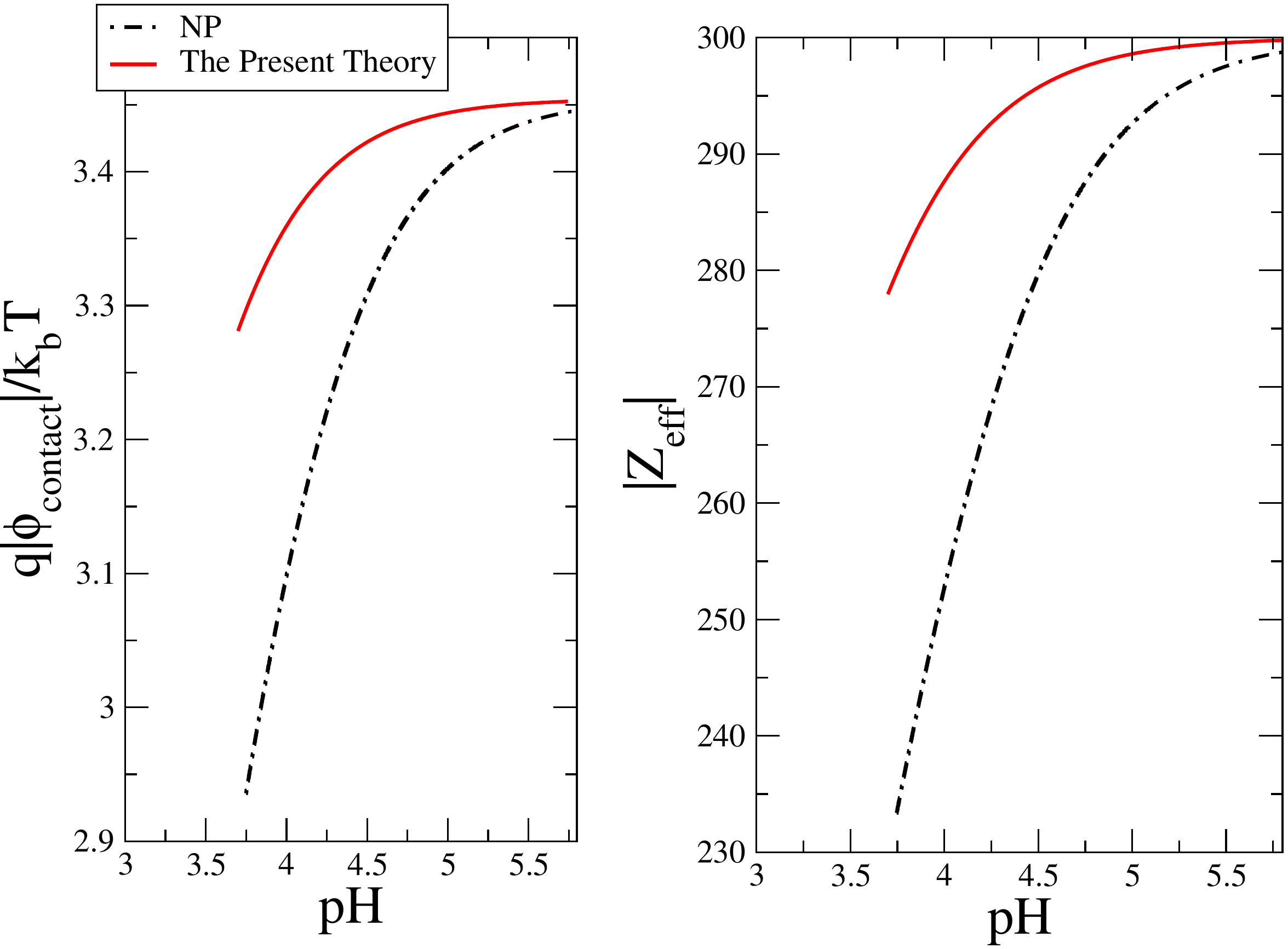}
		\end{center}
		\caption{{ The modulus of the effective charge in units of $q$ and the contact potential of a nanoparticle as a function of pH in the acid reservoir, predicted by the NP and the present theories. The colloidal particle has $300$ charged  functional group on it surface.  The suspension is in a contact with a monovalent salt reservoir at concentration of $10$ mM. The parameters are $a =100$ \AA, $R = 200$ \AA, and $l_g=109.97 $\AA. } 
		}  
		\label{cs}
	\end{figure}

	In Fig.~\ref{fig7} we show the behavior of the effective charge and contact potential of colloidal particle as a function of  1:1 salt concentration for different pH values.
	%%%%%%%%%%%%%%%%%%%%%%%%%%%%%%5
	%%%%%%%%%%%%%%%%fig%%%%%%%%%%%%%%%%%%%
	The figure shows that increase of salt concentration leads to increase of the modulus of the effective charge.  This, again, is a consequence of electrostatic screening produced by salt on the Coulomb interaction between hydronium ions and the negatively charged adsorption sites --- making the association of a hydronium with an active site less energetically favorable. In Fig.~\ref{figadded} we compare the effective charge and contact potential  calculated using the  present theory and the values predicted by the NP theory, for nanoparticles with $300$ charged groups.  As can be seen, neglect of discrete charge effects in the NP theory leads to smaller modulus of the contact potential and of the effective charge. We also note that at large pH the effective charge saturates at the value smaller than the bare charge.  This is a consequence of the overall charge neutrality of the colloidal suspension.  Even if the reservoir has a very small concentration of acid -- large pH, in the absence of other cations inside the suspension, there must be enough hydronium ions to compensate all the colloidal charge.  Some of these hydronium ions will then associate with the surface groups, leading to the saturation of the effective colloidal charge.  
	%%%%%%%%%%%%%%%%%%%%%%%%%%%%%%5
	%%%%%%%%%%%%%%%%fig%%%%%%%%%%%%%%%%%%%
	 We now perform the same calculation, but in the present of a reservoir with $10$ mM monovalent salt. As can be seen in Fig. \ref{cs}, in the presence of salt, for high pH both NP and our theory predict that the effective charge approaches the bare charge.  This is should be contrasted with the no-salt system.  When the system is connected to both the salt and acid reservoirs, at large pH  the hydronium ions inside the system are replaced by the salt cations, which then control the overall charge neutrality of the colloidal suspension.  Since in our model salt cations do not react with the surface groups, for reservoir at large pH very few hydronium ions will be present inside the suspension.  Therefore, all the surface groups will become ionized, and the effective colloidal charge will approach the value of the bare charge. 
	%%%%%%%%%%%%%%%%%%%%%%%%%%%%%%5
	\section{Mixture of functional groups}
	
	As a final example we consider a colloidal particle with a mixture of acidic and basic surface groups.  
	Following the same approach introduced in the previous sections  we find that the effective surface charge is
	\begin{equation}
	\begin{split}  
	\sigma_{eff} = -\frac{N_{acid} q}{4 \pi(a+r_{ion})^2} + q (l^{eff}_{gc}~ c_a~ e^{-\beta \varphi_0^{c} } + l^{eff}_{gn}~c_a~e^{-\beta \varphi_0^{n} } ), \\ 
	l_{gc}^{eff} =\frac{ l_{gc}~N_{acid}~r_{patch}^2}{4  (a+r_{ion})^2\left(1+ l_{gc} c_a \mathrm{e}^{-\beta\varphi_0^c }\pi r_{patch}^2\right)},\\
	l_{gn}^{eff} =\frac{ l_{gn}~N_{base}~r_{patch}^2}{4  (a+r_{ion})^2\left(1+ l_{gn} c_a \mathrm{e}^{-\beta\varphi_0^n }\pi r_{patch}^2\right)},\\
	\end{split}
	\label{Eqadd5}
	\end{equation}
	where $N_{acid}$  is the number of acidic groups and  $N_{base} = N_{site}-N_{acid} $ is the number of basic groups.  The effective sticky length for acidic (charged) and basic (neutral) groups are:  $l^{eff}_{gc}$ and $ l^{eff}_{gn}$, respectively. The discreteness effects will manifest themselves in different ways for hydronium ions condensing on acidic and basic groups, 
	\begin{equation}  
	\begin{split}
	\beta\varphi_0^c = \beta\phi_0^c +\mu^{qq}_c,\\
	\beta\varphi_0^n = \beta\phi_0^n +\mu^{qq}_n,\\
	\end{split}
	\label{Eqadd4}
	\end{equation}
	Since the values $\mu^{qq}_{c,n}$  depend only on the electrostatic  interaction between the hydronium ion and the charged (acid) sites, the value of $\mu^{qq}_c$ will be the same as in Eq. (\ref{v11}), depending  only on the average separation between the {\it acidic} groups.  We will suppose that the basic groups are also uniformly distributed on the colloidal surface on a dual hexagonal lattice with vertexes at the center of each triangle composed of acidic sites.  In this case the position of one of the basic groups will be at $x_0 = d/2 $,$y_0 = \sqrt{3 d}/4$.  Using Eq. (\ref{v10a})  we obtain $\mu^{qq}_n$
	\begin{equation}
	\begin{split}
	\mu_n^{qq} = -\frac{2 \pi q^2}{\gamma \epsilon_w}\sum_{n=-\infty}^{n=\infty \prime} \sum_{m=-\infty}^{m=\infty \prime} \frac{\mathrm{e}^{-2 k r_{ion}}}{k}\cos{\pi \left(m+\frac{n}{2}\right)}
	\end{split}
	\label{v12}
	\end{equation} 
	
	The effective surface charge density can now be written as
	\begin{eqnarray}  
	\sigma_{eff} &=& -\frac{N_{acid}~q}{4 \pi\left(a+r_{ion}\right)^2} +\frac{q K_{Surf}^a~N_{acid}~c_a \mathrm{e}^{-\beta \phi_0}}{4 \pi (a+r_{ion})^2\left(1+ K_{Surf}^a c_a \mathrm{e}^{-\beta \phi_0}\right)}+ \nonumber\\
	&&\frac{q K_{Surf}^b  N_{base}c_a \mathrm{e}^{-\beta \phi_0}}{4 \pi (a+r_{ion})^2\left(1+ K_{Surf}^b c_a \mathrm{e}^{-\beta \phi_0}\right)}
	\label{Eqadd3}
	\end{eqnarray}
	where 
	\begin{equation}
	K_{Surf}^a = \frac{K_{Bulk}^a}{2}\mathrm{e}^{-b-\beta \mu^{qq}_c},
	\label{surfa}
	\end{equation}
	and
	\begin{equation}
	K_{Surf}^b = \frac{K_{Bulk}^b}{2}\mathrm{e}^{-\beta \mu^{qq}_n}.
	\label{surfb}
	\end{equation}
	The bulk equilibrium constants for acid and base, $K_{Bulk}^a$ and $K_{Bulk}^b$,  are given by  Eqs.~(\ref{ka}) and (\ref{Eq3-5}), respectively.
	To test our theory for mixture of basic and acidic surface groups we, once again,  compare it with MC simulations.  
	We consider a colloidal particle with $500$ adsorption sites with  
	different number of acidic groups $N_{acid}$. 
	The sticky lengths --- $l_{gc}$ and $l_{gn}$ --- are $109.97$ and $1099.7$,
	respectively. These values correspond to the equilibrium constants $K_{eq}=$ $0.012$ and $0.00125$ M, respectively.  The concentration of strong acid, HCl, in the reservoir is fixed at $10$ mM. We note that as the value of the sticky length  $l_g$  increases, it becomes progressively more difficult to equilibrate the simulations.   For this reason we have chosen values of $l_g$ that are not too large.  This, however, has no implication for the theory, which remains valid for arbitrary values of  $l_g$ and  $K_{eq}$.
	
	Since the theory is completely general,  the values of $l_g$ are arbitrarily and one can, in practice, choose the depth and the width of the sticky potential and calculated the sticky length.  There is, however, an additional constraint.  The equilibration of the simulations becomes progressively more difficult with increase of sticky length.  To have a good test of the theory we, therefore, need to chose sufficiently large sticky length to have a significant association of hydronium ions with the adsorption sites, while keeping a reasonable equilibration CPU time.  Furthermore, to better test the validity of the theory, we should choose very different sticky lengths for acid and base sites.   This is the reason for a factor of $10$ difference between the values of $l_g$ of acidic and basic groups.  In Table.~\ref{table} we show the values of $\mu^{qq}_c$ and $\mu^{qq}_n$, calculated using Eqs. (\ref{v11}) and (\ref{v12}), respectively  --  for a colloidal particle with the total of $N_{site}=500$ adsorption sites, $N_{acid}$ of which are acidic (charged) and the rest are basic. 
	\begin {table}
	\begin{tabular}{|c|c|c|c|c|c|}
		\hline 
		$N_{acid}$ & 500 & 450 & 300 & 200 & 50 \\ 
		\hline 
		$\mu^{qq}_c$ & -0.6278 & -0.6653 & -0.8073 & -0.94227 & -1.31329 \\ 
		\hline 
		$\mu^{qq}_n$ & --------- & 0.14311 & 0.1519 & 0.1529 & 0.119273 \\ 
		\hline 
	\end{tabular}
	\caption{Different values of $\mu^{qq}_c$ and$\mu^{qq}_n$ for mixture of charged sites. The total number of adsorption sites is $500$} 
	\label{table}  
\end{table}

\begin{figure}
	\begin{center}
		\includegraphics[width=7cm]{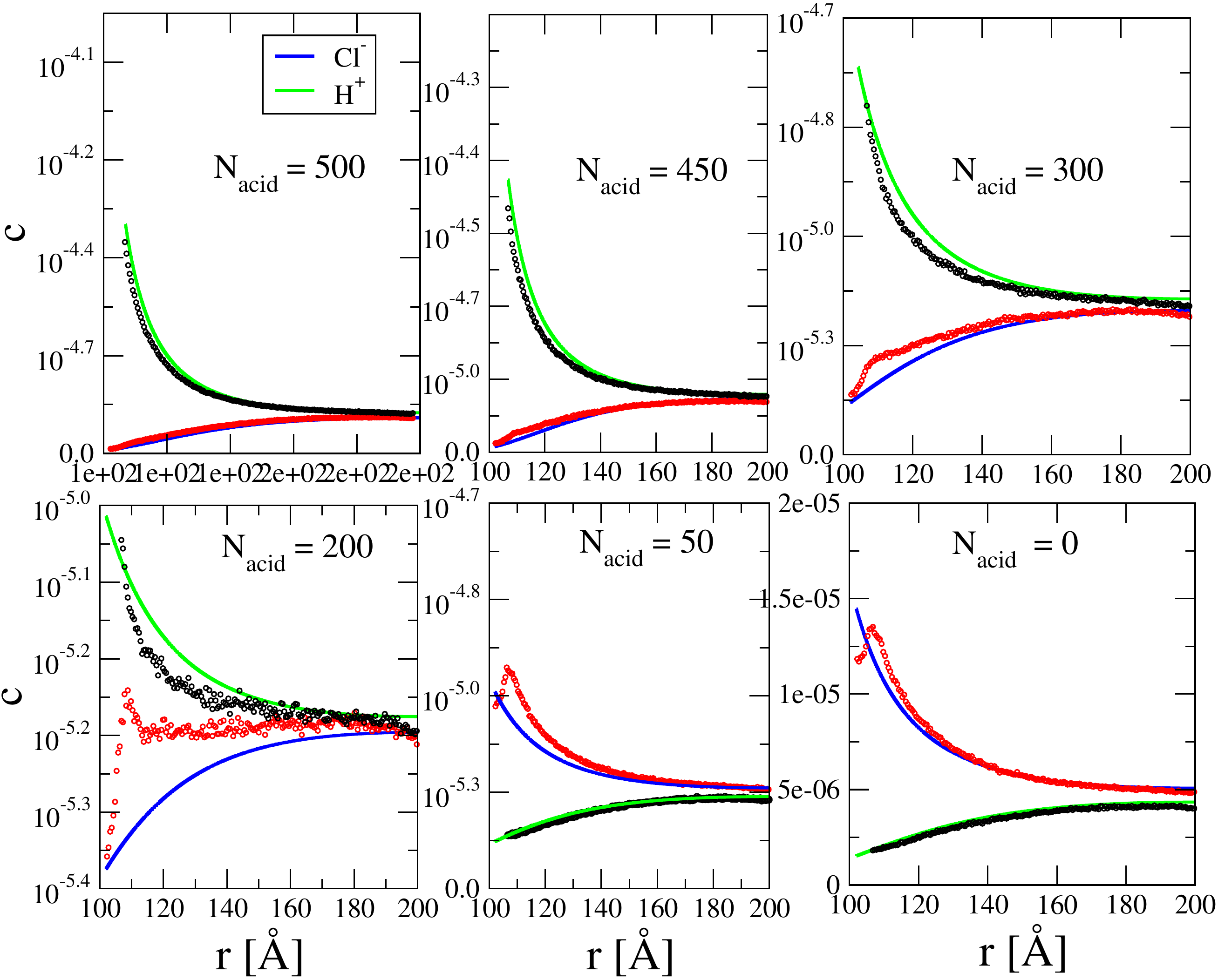}
	\end{center}
	\caption{Density profile of ions around colloidal particle with different number of charged and neutral functional groups. The total number of sites is $N_{site}=500$, the of the reservoir is pH = $2$ and there is no additional salt.  Symbols are the results of MC simulations and the lines are predictions of the theory. The densities are in units of
		particles per \AA$^3$
	}  
	\label{fig_l}
\end{figure}

Fig.~\ref{fig_l} shows that for  small number of basic sites, the theory remains very accurate.  This is also the case  if the number of basic sites  is significantly larger than the number of acidic sites.  The worst agreement is found when $N_{acid}\approx N_{base}$ in which case the surface of colloid becomes strongly heterogeneous, with positive, negative, and neutral domains present, leading to the breakdown of  assumptions used to calculate $\mu^{qq}_{c,n}$. 

From the obtained results, we conclude that the theory works very well if colloidal particle has either basic or acidic adsorption sites.  The discrete charge effects are embedded in the  $\mu^{qq}_c$ and  $\mu^{qq}_n$, which are calculated using a regular arrangement of adsorption sites, even though in the simulations the sites are randomly distributed.  If the number of acid and base sites is approximately equal, then after the adsorption, we will end up with large domains composed of $-1$,$0$,$+1$ charges, and our assumptions for calculating $\mu^{qq}_c$ and  $\mu^{qq}_n$  will break down.  Nevertheless the theory is found to work quite well,  as long as the number of acidic and basic sites is not the same.  Addition of salt to the system results in even better agreement between theory and simulations.  Therefore, the salt free case, provides the most stringent test of the theoretical approach. 
%%%%%%%%%%%%%%%%%%%%%%%%%%%%%%5
%%%%%%%%%%%%%%%%fig%%%%%%%%%%%%%%%%%%%

\section{Conclusion}

In this work, we used a sticky sphere model to mimic chemical reaction on a colloidal 
surface. Within our theory, the discrete charge effects come only from acid surface sites, since ions are treated at the mean field Poisson-Boltzmann level.  This is the reason why the surface equilibrium constant is dependent only on the value of $\mu^{qq}$.  In the previous work we studied colloidal particles with only acidic surface groups~\cite{bakhshandeh2019}.  In that approach we used one component plasma model (OCP), to account for the electrostatic corrections due to discrete surface groups. While the approach in Ref.~\cite{bakhshandeh2019} was sufficiently accurate for colloidal particles with only acidic groups, the OCP does not take into account ionic radius, which prevented us from extending this approach to colloidal particles with a mixture of acidic and basic surface groups.  In the present study, we have developed a completely different method to account for discrete surface effects using periodic Green functions.  The fact that theory works well even for very small nanoparticles of $10$~nm radius shows the robustness of our approach. 
%Sticky potential is particularly suitable for this representation as it avoids ambiguities 
%how to define a bond:  two sticky spheres in direct contact are chemically bonded, if not in direct contact, 
%they are separate molecules.    
%A simple microscopic model of a system undergoing a chemical reaction 

The microscopic model presented in the paper permits us to study the same chemical reaction taking place in bulk and at the interface. In the case of neutral basic surface groups our theory reduces to the NP approach with the bulk equilibrium constant replaced by
the surface equilibrium constant,    
\be
K_{Surf}^b\to \frac{1}{2}K_{Bulk}^b. 
\label{eq:con}
\ee
The difference between surface and bulk equilibrium constants is a consequence of 
steric repulsion, which restricts the overall surface area of the adsorption sites available for  interaction with hydronium ions.

For colloidal particles with acidic surface groups the situation is significantly more complex.  In this case our theory reduces to the NP approach with an effective surface equilibrium constant only for weak acidic groups. For such systems we find the surface equilibrium constant to be 
\begin{equation}  
K_{Surf}^a = \frac{K_{Bulk}^a}{2} \mathrm{e}^{-b-\beta \mu^{qq}_c},
\label{Eqa7}
\end{equation}
where $\mu^{qq}_c$ accounts for the discreteness of surface charge.
For colloidal particles  with a mixture of acidic and basic surface groups, the respective surface equilibrium constants are given by Eqs. (\ref{surfa}) and (\ref{surfb}).

It is important to stress that Eqs. (\ref{eq:con}) and (\ref{Eqa7}) are not universal, and in general will 
depend on the details of the system.  
These details may include molecular geometry, modified electronic structure of surface
functional groups, water structure, etc.  Nevertheless the model of sticky adsorption sites  demonstrates that there is a
mapping between the bulk and the surface equilibrium constants which allows one to use the Poisson-Boltzmann framework
to accurately account for the charge regulation in colloidal systems.  Any deviations from experiment can therefore be attributed to the shortfall of the model and not to the theoretical method used to solve it.

In this work our primary goal was to explore the extent of  validity of the mean-field NP approach by applying it to an exactly solvable model.  Clearly the microscopic model that we used for  spherically  symmetric  hydronium,  uniform  dielectric  water,  sticky  interactions  for covalent binding, etc., is a very rough approximation to the physical reality.  The advantage is that we can solve this model exactly using computer simulations.  Applying the NP approach to the same model we can then test the extent of validity of the mean-field approximations. We should stress that the NP theory does not give us any information whatsoever about the surface equilibrium constant and assumes it to be the same as the bulk association constant. We find, on the there hand,  that sticky interactions result in a breakdown of the mean-field approximations. Surprisingly, however, we find that all the discreteness effects can be included in a renormalized surface association constant, which our theory predicts explicitly.
For our microscopic model the correlations  and  steric  effects  lead  to  lower  surface  association constant $K_{Surf}$,  compared  to  the  bulk association constant for the same acid or base, $K_{Bulk}$.   This means that  fewer  hydronium  ions will bind  to surface groups, implying that  surface p$K_a$ will be smaller than bulk p$K_a$.  Within the present model, there are two contributions which account for the decrease of the association constant at the surface.   First,  is  the  steric  repulsion  from  the  colloidal  surface,  which  diminishes  the  access  of hydronium to  acid and base groups.  Within our model the accessible area  for  the  charge  transfer  reaction  is  lowered  by  a  factor  of  two,  which  accounts  for  the factor  of  $1/2$  which  appears  in  the  surface  binding  constant.   The  second  contribution  comes from the discrete nature of surface charged groups, which we also find to lower the effective binding constant.  On the other hand, the experiments indicate that the surface binding constant,  $K_{Surf}$,  that one needs to use in the NP theory is actually larger than $K_{Bulk}$.  This means that the surface p$K_a$ is larger than the p$K_a$ of the bulk acid~\cite{Behrens}.  Since our model already takes into account all the steric and electrostatic effects at the dielectric continuum level,  we must conclude that in order to account for the experimental results we must included additional effects into the model, such dielectric discontinuity across the colloidal structure, water ordering, quantum nature of proton transfer, etc. The approach that we have developed should allow us to explore these additional effects in order to understand the mechanisms that lead to the increase of  p$K_a$ at colloidal surface. This will be the subject of the future work.

%%%%%%%%%%%%% end of figure %%%%%%%%%%%%%%%%%
\section{ACKNOWLEDGMENT}
This work was partially supported by the CNPq.
%%%%%%%%%%%%%%%% figure %%%%%%%%%%%%%%%%%%%%%
%\begin{figure}
%\begin{center}
%\includegraphics[width=7cm]{zgraph1.eps}
%\end{center}
%\caption{Density profile of positive and negative ions near the charged plate for the case AII. }  
%\label{fig5}
%\end{figure}
%%%%%%%%%%%%% end of figure %%%%%%%%%%%%%%%%%
%%%%%%%%%%%%%%%% figure %%%%%%%%%%%%%%%%%%%%%

%%%%%%%%%%%%% end of figure %%%%%%%%%%%%%%%%%
\bibliographystyle{ieeetr}
\bibliography{ref}
%%%%%%%%%%%%%%%%%%%%%%%%%%%%%%5
%%%%%%%%%%%%%%%%fig%%%%%%%%%%%%%%%%%%%
\end{document}